%% file: main.tex
\documentclass[final,5p,times,twocolumn]{elsarticle}

\usepackage[T1]{fontenc}
\usepackage{amsmath,amssymb,amsthm}
\usepackage{graphicx}
\usepackage{booktabs}
\usepackage{array}
\usepackage{tabularx}
\usepackage{xurl}
\usepackage{siunitx}
\usepackage{algorithm}   
\usepackage{algpseudocode}
\usepackage{pdfpages}
\usepackage[colorlinks=true,linkcolor=blue!50!black,citecolor=blue!50!black,urlcolor=blue!50!black]{hyperref}
\usepackage[capitalise,noabbrev]{cleveref}
\usepackage{xcolor}
\makeatletter\makeatother   

\usepackage{stfloats}   

\graphicspath{{figures/}}
\newtheorem{proposition}{Proposition}
\newtheorem{remark}{Remark}

\newcommand{\R}{\mathcal{R}}
\newcommand{\F}{\mathcal{F}}
\newcommand{\dd}{\mathrm{d}}

\journal{Computer Methods in Applied Mechanics and Engineering}

\begin{document}

\begin{frontmatter}

\title{Error-controlled switching between discrete and continuum dislocations in a unified defect framework}

\author{Albert Linda\corref{cor1}}
\ead{albert.hzbn@gmail.com}
\cortext[cor1]{Corresponding author.}
\affiliation{organization={Department of Materials Science and Engineering, Indian Institute of Technology Kanpur},
            city={Kanpur},
            postcode={208016},
            country={India}}

\input{sections/abstract}


\end{frontmatter}

\input{sections/introduction}
\input{sections/contract}
\input{sections/bounds}
\input{sections/methods}
\input{sections/results_models}
\input{sections/fixed_representation}
\input{sections/allocator}
\input{sections/validation}

\input{sections/copper}
\input{sections/alloc3d}
\input{sections/creep}
\input{sections/discussion}
\input{sections/conclusions}

\input{sections/declarations}

\bibliographystyle{elsarticle-num}
\bibliography{references}

\clearpage
\includepdf[pages=-,pagecommand={\thispagestyle{empty}}]{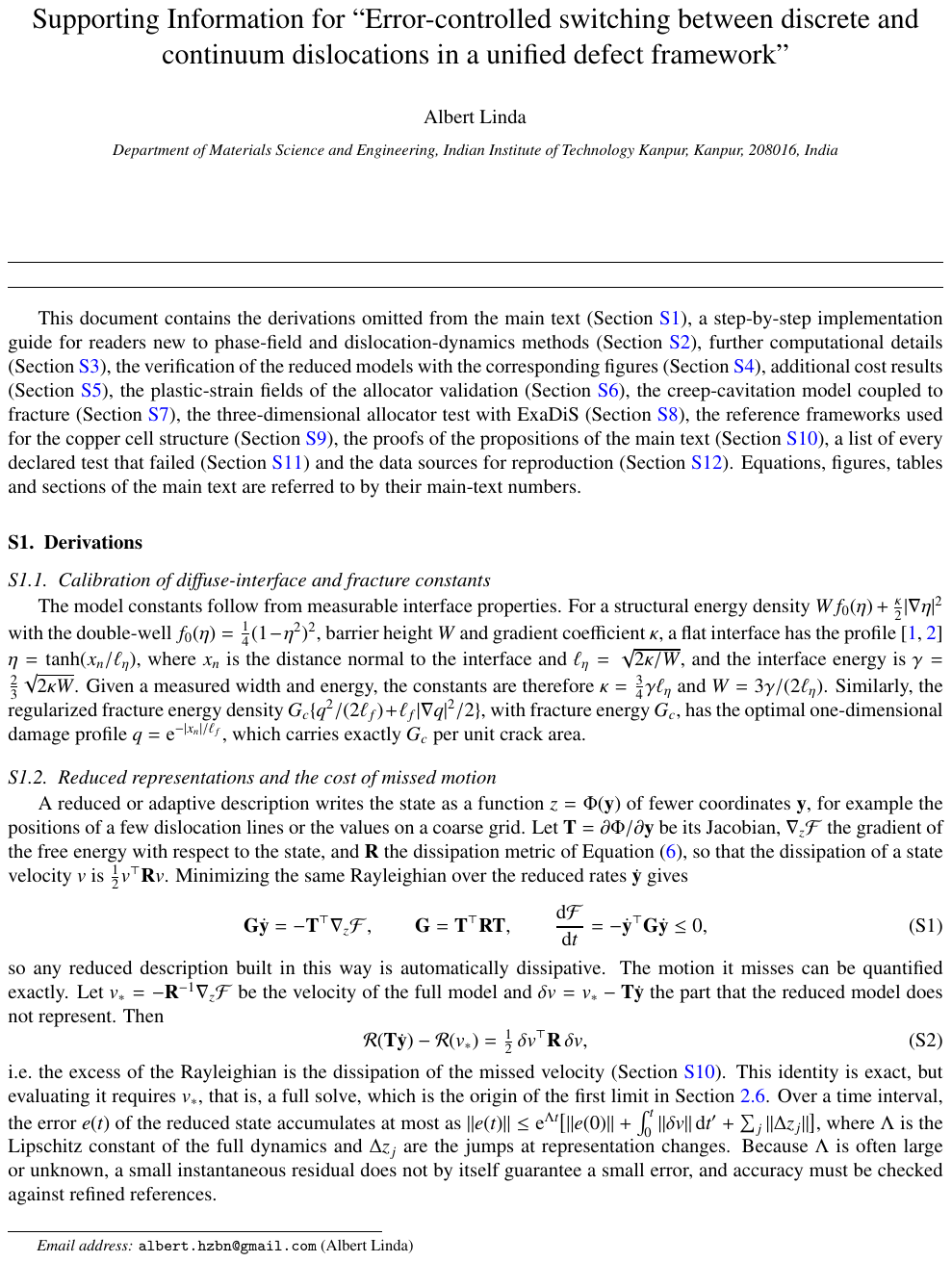}

\end{document}

%% file: sections/abstract.tex
\begin{abstract}
Simulating defects of every dimension together is expensive, and the savings of adaptive methods are rarely tested at equal accuracy. We present a variational framework in which fields, dislocation lines and defect densities share one free energy, one dissipation law and exact ledgers of mass and Burgers vector, so that a region can change its representation without gaining or losing defects. On this basis, an allocator switches the dislocations themselves between discrete particles, where they are sparse, and a continuous density, where they are dense. The switch is made block by block from local counts, conserves the Burgers vector exactly and is controlled by a single threshold. For $5\times10^5$ dislocations, the allocator reproduces the discrete model to within 1\% in the sparse region at 14--37 times lower cost, depending on the machine; the advantage grows to 81 times at $10^6$ dislocations, and a threshold calibrated on one ensemble meets its error target on an independent one. For an idealized array of dislocation walls in copper, the allocator is 37.5 times cheaper than particle dislocation dynamics and, at almost the same cost, six times more accurate than continuum dislocation dynamics between the walls. In a three-dimensional copper model coupling grain-boundary cavitation with damage, ligaments left at nominal coalescence delay creep rupture by 25--48\% (post hoc sweep). By contrast, grids refined at interfaces gain less than twofold once the active region exceeds about 30\% of the domain, so representation switching, rather than grid refinement, is the effective form of adaptivity for defect-rich microstructures.
\end{abstract}


%% file: sections/introduction.tex
\section{Introduction}\label{sec:intro}

The properties of metals and alloys are governed by defects of every dimension. Vacancies and solute atoms are point defects, dislocations are line defects, grain and phase boundaries and crack faces are planar defects, and precipitates, pores and grains are volume defects. These defects rarely act alone. Vacancies diffuse to dislocations and pores, dislocations pile up at grain boundaries, and cracks deflect at tough particles. Predicting how a microstructure evolves therefore requires following defects of different dimensions, and the interactions between them, within a single calculation.

The phase-field (PF) method is the most widely used tool for this purpose~\cite{chen2002,moelans2008,tourret2022}. It represents each interface as a smooth field of finite width and derives the evolution from a free energy and a dissipation law~\cite{cahn1958,allen1979}. The same framework has been applied to multiphase and polycrystalline microstructures~\cite{steinbach1996}, dislocation cores and slip~\cite{koslowski2002,wang2001}, solute segregation at dislocations~\cite{hu2001} and fracture~\cite{francfort1998,bourdin2000,miehe2010ijnme,miehe2010cmame,borden2012}, including ductile fracture coupled to plasticity~\cite{miehe2016ductile}. PF simulations are expensive, however, and computational cost remains one of the main challenges identified in recent reviews~\cite{tourret2022}. Every interface must be resolved by several grid points, and a conventional implementation solves every field over the whole domain, including bulk regions in which nothing happens. Moreover, defects of different dimension are usually treated by separate methods. Point defects are described by rate theory, dislocations by discrete~\cite{vdg1995,zbib1998,arsenlis2007,cai2006} or continuum~\cite{groma1997,acharya2001,hochrainer2014,xia2015} dislocation dynamics, and surfaces by PF, level-set~\cite{osher1988} or cohesive models. Joining such methods raises the question of how to avoid counting the same energy twice and how to transfer information consistently between them.

Adaptivity is the usual remedy for the cost. One approach keeps the equations unchanged and refines the grid only near interfaces, as in adaptive mesh refinement (AMR)~\cite{berger1989,provatas1999,feng2008}, now also on GPUs~\cite{sakane2022}, adaptive amplitude expansions~\cite{athreya2007} and adaptive reduced models~\cite{grassle2017}; mature AMR codes are available~\cite{amrex2019,prismspf2020,moose2020}. A second approach changes the representation, replacing many discrete objects by a density, a resolved field by a line, or fine grid detail by an enrichment, as in partition-of-unity methods~\cite{melenk1996} and the quasicontinuum method~\cite{tadmor1996}. Data-driven surrogates learned from dislocation-dynamics data~\cite{salmenjoki2018} offer a third route, but they must be trained on the full model and carry no built-in conservation of mass or Burgers vector. The reported gains of all these approaches vary widely between problems. Grid refinement pays off when the active region is small, as for isolated dendrites~\cite{sakane2022}, but not necessarily when defects fill a large part of the domain, and comparisons at equal physical interface width and equal accuracy are uncommon, although widening an interface to save work changes the model being solved.

For dislocations, the two classical descriptions lie at opposite ends of this trade-off. Discrete dislocation dynamics tracks every line accurately, but its cost grows with the number of dislocations~\cite{vdg1995,arsenlis2007}; even with GPU acceleration and subcycling~\cite{bertin2019}, dense and heavily strained microstructures remain a frontier of the field~\cite{bertin2020}. Continuum dislocation dynamics evolves densities at a cost set by the grid~\cite{groma1997,hochrainer2014,xia2015}, and recent work has added annihilation, sources and junction reactions~\cite{monavari2018,linelazab2020} and robust numerical schemes~\cite{lin2021}. A density, however, cannot represent the individual dislocations that dominate sparse regions, and direct comparisons show that the two descriptions differ in their microstructural kinematics~\cite{sandfeld2015}. Existing couplings keep the two roles fixed. The discrete--continuum model~\cite{lemarchand2001,vattre2014}, in its consistent recent formulation~\cite{jamond2016}, and related multiscale models~\cite{zbib2002} pass the plastic strain of discrete dislocations to a finite-element continuum while every dislocation remains discrete. Wallin et al.~\cite{wallin2008} hand discrete dislocations across a fixed interface to a crystal-plasticity region, in one direction only, and a dislocation-density field has been coupled to discrete dislocation dynamics in time rather than in space~\cite{chatterjee2020}. Crystal-plasticity frameworks have recently been coupled to continuum dislocation dynamics~\cite{yang2025}, in which every dislocation is a density. Concurrent atomistic--continuum methods~\cite{tadmor1996,xu2017cac} bridge atomic and continuum scales around individual dislocation cores, but do not address the collective transition between discrete lines and densities. Adaptive switching between particle and continuum descriptions is established for rarefied gases~\cite{garcia1999,schwartzentruber2007}, molecular liquids~\cite{praprotnik2005} and granular media~\cite{hybridgrains2018}. We are not aware of a scheme in which the dislocations themselves switch, in both directions and as the local density changes, between discrete particles and a transported dislocation density, where the Burgers vector must be conserved exactly and the elastic energy must not be counted twice. In general, adaptive defect simulations rarely specify together which energy each description carries, what is conserved when a region changes description, and how the error of the switch is controlled and at what cost.

In this study, we develop a variational framework in which phase and composition fields, dislocation lines and defect densities share one free energy, one dissipation principle and exact ledgers of mass and Burgers vector, so that a region can change its representation without creating or losing defects, and any energy change at a switch is evaluated with the same free energy. On this basis we construct an error-controlled allocator that treats dislocations as discrete particles where they are sparse and as a transported density where they are dense. The switching decision is taken in advance from local dislocation counts, each conversion conserves the Burgers vector exactly, and a single threshold sets the accuracy. We first verify the framework against established models of interface motion, coarsening, grain growth, precipitation, point-defect transport, pore evolution, dislocation mechanics and fracture, and calibrate its closures for copper with atomistic calculations and published data. We then derive three elementary results on the cost of adaptivity and compare interface-focused grids with uniform grids and an independent AMR code at equal accuracy. Finally, we measure the allocator against discrete and continuum dislocation dynamics, and validate it on a second machine against a reference refined fourfold in space and time (converged to 0.2\% overall but only to about 7\% in the sparse region), over ensemble sizes spanning more than two orders of magnitude and five independent ensembles. The contribution of this work is threefold. First, a single energy, dissipation and ledger framework is formulated under which fields, dislocation lines and defect densities can be exchanged locally without double counting. Second, the dislocation representation itself is switched adaptively, block by block, with an a priori criterion, exact Burgers-vector conservation and one error-controlling threshold whose calibration transfers between ensembles. Third, simple provable limits and fair, equal-accuracy measurements establish when adaptivity reduces cost and when it cannot. The individual ingredients (diffuse interfaces with local resolution, conservative transfer, regularized dislocation lines, variational fracture and particle--continuum partitioning in other fields) have precedents, which we cite where they are used. Except for the three-dimensional dislocation calculations with the production code ExaDiS (C++/Kokkos) and the GPU creep model (CuPy), all methods compared here are implemented in Python on single computers. Most calculations are two-dimensional with dimensionless parameters. Every test was defined with its acceptance criteria before it was run, failed tests are reported as failed, and analyses added after a result had been seen are labelled post hoc.

%% file: sections/contract.tex
\section{Theoretical framework}\label{sec:contract}

The framework rests on three requirements. All representations share one free energy, so that no energy is counted twice. All evolve by one dissipation principle, so that the free energy of an isolated system cannot increase. Every change of representation keeps exact ledgers of the conserved quantities, such as mass and Burgers vector.

\begin{figure*}[!tb]
\centering
\includegraphics[width=\textwidth]{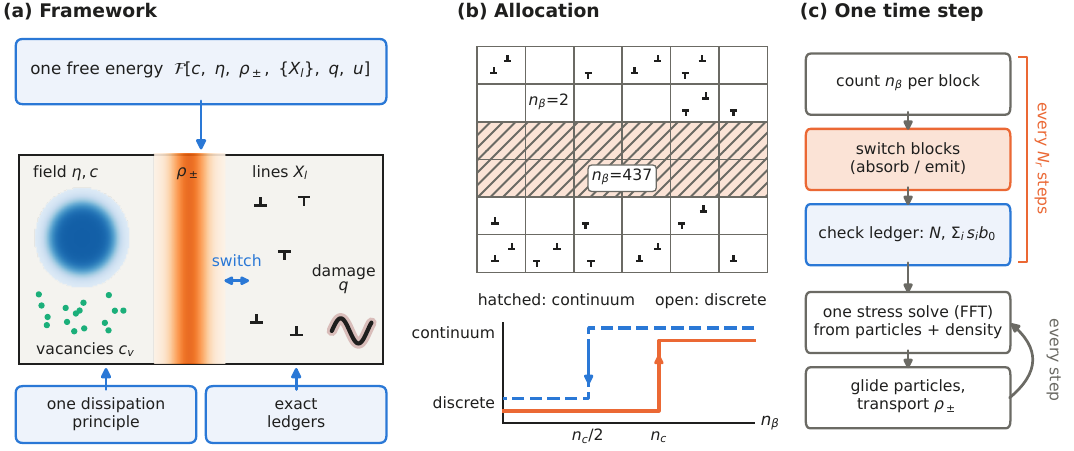}
\caption{Schematic of the framework and the allocator. (a) Phase and composition fields ($\eta$, $c$), dislocation densities ($\rho_\pm$), dislocation lines ($X_l$), damage ($q$) and point defects share one free energy $\mathcal F$, evolve by minimising one Rayleighian $\mathcal R$, and keep exact ledgers of mass and Burgers vector. (b) The allocator divides the domain into blocks $\beta$; a block becomes continuum (hatched) when its dislocation count $n_\beta$ reaches the threshold $n_c$, and returns to discrete only when $n_\beta$ falls below $n_c/2$. Sparse blocks keep individual dislocations. (c) The two conversions: particles are absorbed into $\rho_\pm$ with the cloud-in-cell weights used for the stress, and densities emit $\lfloor\rho_\pm h^2\rfloor$ particles per cell of size $h$, keeping the fractional remainder; both leave the number $N$ of dislocations and their net Burgers vector $\sum_i s_ib_0$ unchanged.}\label{fig:schematic}
\end{figure*}

A simple picture helps to read the equations that follow. The free energy is a landscape over all possible states of the material, and the microstructure is a heavily damped ball rolling downhill in it: it always moves towards lower energy, and friction (dissipation) sets how fast. Vacancies diffusing to a sink, a grain boundary straightening, a dislocation gliding under stress and a crack opening are all this same downhill motion, in different directions of the landscape and with different friction. The framework writes one landscape for all defects together, so that their interactions are automatic, and one friction law for all motions, so that no process can create energy. The ledgers are bookkeeping on top of this: when the description of a region changes, for example from individual dislocations to a density, the totals must stay the same. \Cref{fig:schematic} summarises these requirements and the allocator built on them. \Cref{tab:notation} collects the symbols used in this section and in \cref{sec:methods}; each symbol is also defined where it first appears.

\begin{figure*}[!b]
\centering
\includegraphics[width=\textwidth]{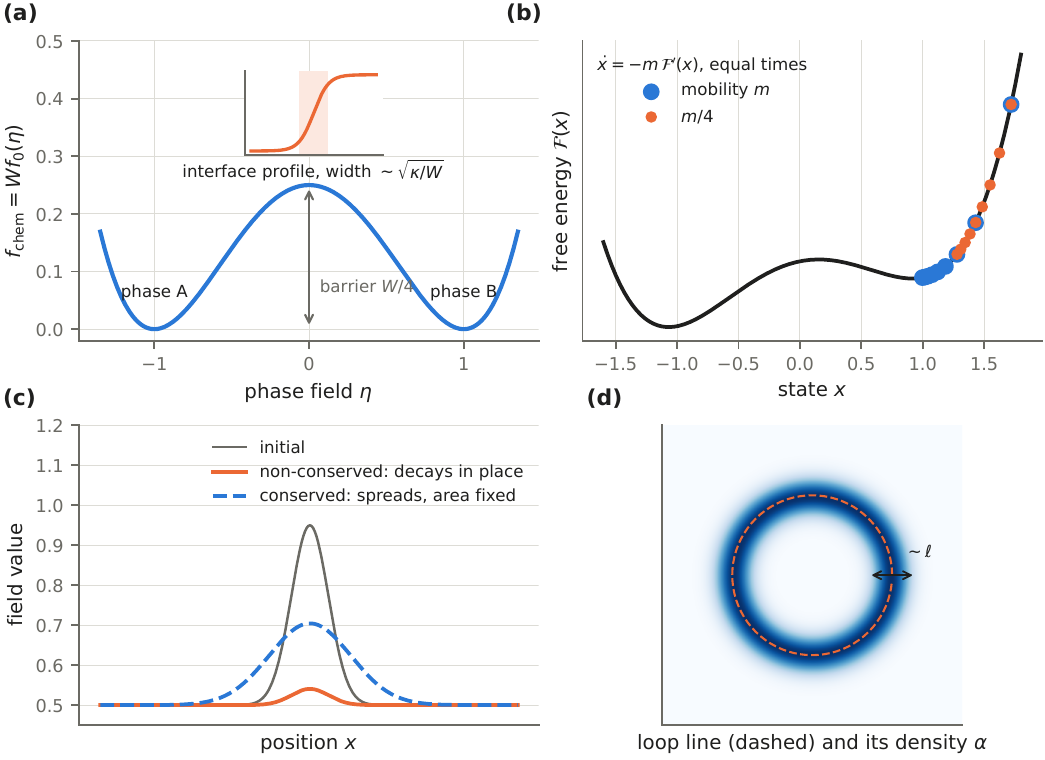}
\caption{The ingredients of the framework in one dimension. (a) Double-well energy $Wf_0(\eta)$ with its two minima (the two phases) and barrier; the balance with the gradient energy gives a smooth interface profile of width $\sim\sqrt{\kappa/W}$ (inset). (b) The variational principle for one variable: minimizing the Rayleighian gives $\dot x=-m\mathcal F'(x)$, so the state moves downhill in the energy landscape, faster for a larger mobility $m$ (positions at equal time intervals). (c) Non-conserved relaxation ($\dot\eta=-Lg$) removes an excess in place, whereas conserved relaxation ($\dot c=\nabla\cdot M\nabla\mu$) spreads it while keeping its total fixed. (d) A dislocation loop (dashed) spread over a width $\ell$ becomes a smooth density $\alpha$ that can interact with the fields.}\label{fig:theory}
\end{figure*}

\begin{table*}[!t]
\centering
\caption{Nomenclature. Units are given for dimensional quantities; the benchmarks use dimensionless values.}\label{tab:notation}
\footnotesize
\setlength{\tabcolsep}{4pt}
\begin{tabular}{@{}>{$}l<{$}p{0.36\linewidth}>{$}l<{$}p{0.36\linewidth}@{}}
\toprule
\multicolumn{1}{@{}l}{Symbol} & Meaning & \multicolumn{1}{l}{Symbol} & Meaning \\
\midrule
\Omega,\ d & simulated region, spatial dimension (2 or 3) & \mathcal F,\ \mathcal R & free energy (J), Rayleighian (W) \\
x,\ t & position (m), time (s) & f_{(\cdot)} & free-energy densities (J\,m$^{-3}$) \\
c & composition (site fraction) & \mu=\delta\mathcal F/\delta c & chemical potential (J\,m$^{-3}$) \\
\eta & phase or grain fraction & g=\delta\mathcal F/\delta\eta & structural driving force (J\,m$^{-3}$) \\
q\in[0,1] & damage (0 intact, 1 broken) & \sigma & Cauchy stress (Pa) \\
u,\ \beta^p & displacement (m), plastic distortion & J & composition flux (m\,s$^{-1}$) \\
T & temperature (K) & r,\ S,\ \mathcal A & reaction rates, stoichiometry, affinity \\
X_l,\ b_l,\ \hat{\mathbf t}_l & position, Burgers vector and unit tangent of line $l$ & M,\ K,\ L,\ L_q & diffusion, reaction, phase and damage mobilities \\
s & arclength along a line (m) & B_l,\ P_l & drag tensor, projector onto allowed motion of line $l$ \\
 & & f_l & total force per length on line $l$ (N\,m$^{-1}$) \\
\alpha & Nye tensor (m$^{-1}$) & W,\ \kappa & barrier height (J\,m$^{-3}$), gradient coefficient (J\,m$^{-1}$) \\
\delta_\ell & smoothing kernel of width $\ell$ & \gamma,\ G_c & interface energy, fracture energy (J\,m$^{-2}$) \\
w & physical interface width (m) & \mathbf y & coordinates of a reduced representation \\
a & dislocation core width (m) & \Phi,\ \mathbf T & map from $\mathbf y$ to the state, its Jacobian $\partial\Phi/\partial\mathbf y$ \\
\ell_f & fracture regularization length (m) & \mathbf R & dissipation metric \\
h,\ L_c & fine and coarse grid spacing (m) & v_*,\ \delta v & full-model velocity, part missed by the reduction \\
L_{\mathrm{int}} & interaction range (m) & \Lambda & Lipschitz constant of the full dynamics (s$^{-1}$) \\
C_{(\cdot)} & computational cost of one step (s) & \chi & ratio of residual to full-step cost \\
n_c & allocator threshold (dislocations per block) & D,\ C,\ H,\ D^* & discrete, continuum, allocator; finest discrete reference \\
\bottomrule
\end{tabular}
\end{table*}

\subsection{State variables and length scales}

Let $\Omega\subset\mathbb{R}^d$ be the simulated region, with $d=2$ or $3$ the spatial dimension, $x\in\Omega$ the position and $t$ the time. The state of the material is the collection
\begin{equation}
z=(c,\eta,u,\beta^p,q,T,X),
\end{equation}
in which $c(x,t)$ is the composition, $\eta(x,t)$ the phase or grain fractions, $u(x,t)$ the displacement, $\beta^p(x,t)$ the plastic distortion, $q(x,t)\in[0,1]$ the damage, $T(x,t)$ the temperature, $X$ the positions of explicitly tracked defects (dislocation lines $X_l$ with Burgers vectors $b_l$, clusters with their positions and sizes) and their internal states, such as the size of a cluster. A given calculation uses only the variables its physics requires.

The variables are of two kinds. \emph{Fields} have a value at every point: the composition $c$ is the local fraction of solute or vacancies; the phase field $\eta$~\cite{chen2002,provatas1999} is a smooth indicator that equals one inside a precipitate or grain and zero outside (with the symmetric double well of \cref{S-si:derivations}, $\eta$ runs from $-1$ to $1$ and $(1+\eta)/2$ has this meaning), changing continuously across an interface of finite width; and the damage $q$ is a smooth indicator of broken material that equals zero where the solid is intact and one inside a crack. Interfaces and cracks are therefore not surfaces that must be tracked and remeshed; they are the thin regions where these fields change, and their motion is simply the change of the fields in time. \emph{Tracked defects} $X$, such as dislocation lines, are kept as individual objects with positions, because their discreteness matters where they are few.

To couple a defect of lower dimension to the fields, the defect is spread over a small, fixed width $\ell$, much as a point charge is represented by a narrow charge cloud (\cref{fig:theory}d). Let the defect occupy a set $M_k$ of dimension $k$ (a point for $k=0$, a line for $k=1$ or a surface for $k=2$), parametrized by $\xi$, and let $\rho_k(\xi)$ be a quantity it carries per unit of length, area or count. Its field representation is
\begin{equation}\label{eq:embed}
E_\ell[\rho_k](x)=\int_{M_k}\rho_k(\xi)\,\delta_\ell\bigl(x-X(\xi)\bigr)\,\dd s_k(\xi),
\end{equation}
where $\delta_\ell$ is a smooth kernel of width $\ell$ with unit integral, $\int_{\mathbb R^d}\delta_\ell(x)\,\dd x=1$, and $\dd s_k$ is the element of length, area or count on $M_k$. If $\rho_k$ has units $U$, then $E_\ell[\rho_k]$ has units $U\,\mathrm{m}^{k-d}$, i.e.\ it is a density in the $d$-dimensional region. For example, a closed dislocation line with Burgers vector $b$ and unit tangent $\hat{\mathbf t}(s)$, where $s$ is the arclength, becomes the Nye tensor (dislocation density)~\cite{nye1953,acharya2001}
\begin{equation}
\alpha(x)=\oint b\otimes\hat{\mathbf t}(s)\,\delta_\ell\bigl(x-X(s)\bigr)\,\dd s ,
\end{equation}
which satisfies $\operatorname{div}\alpha=0$ exactly because the line is closed. The dimension of a defect is thus carried by the set on which it lives and by the units of its density, rather than by forcing all defects into one scalar field.

Five kinds of length are kept distinct: (i) the physical widths of the model, namely the interface width $w$, the dislocation core width $a$ and the fracture regularization length $\ell_f$; (ii) the numerical regularization width $\ell$ of the smoothing kernel $\delta_\ell$ in \cref{eq:embed}, which spreads a line or particle over the grid (in the allocator studies $\ell=1.5h$ on the $256^2$ grid, held at this physical value when the grid is refined); (iii) the resolution, given by the fine grid spacing $h$ and the coarse grid spacing $L_c$; (iv) the coarse-graining length over which dislocation counts and densities are defined, which is the size of an allocation block; and (v) the interaction range $L_{\mathrm{int}}$ (for example the range of the elastic field). The lengths in (i) and (ii) belong to the model and are fixed in every comparison; only $h$ and $L_c$ change when a grid is refined. A calculation is therefore never made cheaper by widening an interface, which would change the model.

\subsection{Free energy}

The free energy $\mathcal F$ (J) is the sum of bulk contributions, localized defect contributions and a constraint term:
\begin{equation}\label{eq:F}
\begin{aligned}
\F={}&\int_\Omega\bigl(f_{\mathrm{chem}}+f_{\mathrm{grad}}+f_{\mathrm{el}}+f_{\mathrm{hard}}\\
&\qquad+f_{\mathrm{frac}}+f_{\mathrm{th}}+f_{\mathrm{ext}}\bigr)\,\dd V\\
&+\F_{\mathrm{core}}+\F_{\mathrm{bind}}+\F_{\mathrm{con}}.
\end{aligned}
\end{equation}
Here $f_{\mathrm{chem}}$, $f_{\mathrm{grad}}$, $f_{\mathrm{el}}$, $f_{\mathrm{hard}}$, $f_{\mathrm{frac}}$, $f_{\mathrm{th}}$ and $f_{\mathrm{ext}}$ are the chemical, gradient (interfacial), elastic, hardening, fracture, thermal and external-load energy densities (J\,m$^{-3}$), $\dd V$ is the volume element, $\F_{\mathrm{core}}$ is the core energy of the tracked defects and $\F_{\mathrm{bind}}$ their binding energy to solute and interfaces. The constraint term $\F_{\mathrm{con}}$ is zero for admissible states and infinite otherwise; it enforces the bounds of $c$, $\eta$ and $q$, the requirement that phase fractions sum to one, the balance of Burgers vectors at junctions and the irreversibility of damage.

Each term has a simple meaning. For a single phase field the first three take the standard forms
\begin{gather}\label{eq:terms}
f_{\mathrm{chem}}=W f_0(\eta),\qquad f_{\mathrm{grad}}=\tfrac{\kappa}{2}|\nabla\eta|^2,\\
f_{\mathrm{el}}=\tfrac12\bigl(\varepsilon-\varepsilon^*\bigr)\!:\!\mathbb C\!:\!\bigl(\varepsilon-\varepsilon^*\bigr),\nonumber
\end{gather}
where, following the standard phase-field constructions~\cite{cahn1958,allen1979,chen2002}, $f_0(\eta)$ is a double well with minima at the two phases and barrier height $W$, $\kappa$ is the gradient coefficient, $\varepsilon$ the total strain, $\varepsilon^*$ the eigenstrain carried by the defects (the misfit of a precipitate or the plastic distortion $\beta^p$ of dislocations) and $\mathbb C$ the elastic stiffness. The double well favours the two pure phases, the gradient term penalizes sharp changes, and their balance gives an interface of finite width $w\propto\sqrt{\kappa/W}$ and energy $\gamma\propto\sqrt{\kappa W}$. The elastic term is the energy stored when defects force the lattice away from its stress-free shape, in the eigenstrain formulation of micromechanics~\cite{mura1987,khachaturyan1983}; it is through this single term that a precipitate, a dislocation and a crack feel each other's stress. The fracture energy, $f_{\mathrm{frac}}=\tfrac{G_c}{2}\bigl(q^2/\ell_f+\ell_f|\nabla q|^2\bigr)$, is built in the same way: a crack of area $A$ costs $G_cA$, spread over the length $\ell_f$~\cite{bourdin2000,miehe2010cmame}, and damage reduces the stiffness in $f_{\mathrm{el}}$ by the factor $(1-q)^2$. \Cref{fig:theory}a shows the double well and the resulting interface profile.
Where two descriptions overlap, three rules keep each energy counted once. (i) A dislocation contributes its long-range elastic energy either through the plastic distortion $\beta^p$ and one global elastic solve, or through a closed-form interaction energy of its position, never both; a smoothed (non-singular) core~\cite{cai2006} that already carries the elastic energy $E^{\mathrm{reg}}_{\mathrm{el}}(a)$ per unit length receives only the missing part $E_{\mathrm{target}}-E^{\mathrm{reg}}_{\mathrm{el}}(a)$, where $E_{\mathrm{target}}$ is the calibrated core energy. (ii) Slip carried by explicit lines and slip carried by a smeared crystal-plasticity field occupy separate parts of the state. (iii) Where two diffuse regions overlap, one field on their union carries one energy. Plastic work is split into stored energy and heat, and only the stored part enters $\F$.

The two constants of each diffuse interface, and the fracture length, follow from its measured width and energy (\cref{S-si:derivations}), so that a physical width fixed for a comparison is never adjusted to save cost.

\subsection{Evolution equations}

All evolution follows from one variational principle: the rates minimize the Rayleighian $\R$, which is the dissipation rate plus the rate of change of the free energy. This is Onsager's variational principle~\cite{onsager1931,doi2015}. The idea is clearest for a single variable $x$ with energy $\F(x)$ and friction coefficient $1/m$ (mobility $m$). Moving at rate $\dot x$ costs the dissipation $\tfrac12\dot x^2/m$ and releases energy at the rate $-\dot\F=-\F'(x)\dot x$. The Rayleighian $\R=\tfrac12\dot x^2/m+\F'(x)\dot x$ is smallest for $\dot x=-m\F'(x)$: the variable moves down the energy slope, at a speed proportional to the slope and to the mobility. Every law below is this rule applied to a different variable, with the appropriate friction (\cref{fig:theory}b). Let $J$ be the composition flux, $r$ the reaction rates with stoichiometric matrix $S$, $\dot\eta=\partial\eta/\partial t$ the phase rates and $\dot X_l$ the velocity of line $l$. Then
\begin{equation}\label{eq:rayleighian}
\begin{aligned}
\R={}&\tfrac12\int_\Omega\bigl(J^{\top}M^{-1}J+r^{\top}K^{-1}r+\dot\eta^{\top}L^{-1}\dot\eta\bigr)\,\dd V\\
&+\tfrac12\sum_l\int\dot X_l\cdot B_l\dot X_l\,\dd s+\dot\F,
\end{aligned}
\end{equation}
subject to the mass balance $\dot c=-\nabla\cdot J+Sr$. Plastic flow and damage enter $\R$ in the same way, through convex dissipation potentials $\phi_{\mathrm{pl}}(\dot\beta^p)$ and $\phi_q(\dot q)$, the latter with the constraint $\dot q\ge0$; they are omitted from \cref{eq:rayleighian} for brevity. The first three terms are the dissipation of diffusion, reactions and phase change, with positive mobilities $M$, $K$ and $L$; the fourth is the dissipation of dislocation motion, with a positive drag tensor $B_l$; and $\dot\F$ is the rate of change of the free energy. Setting the variation of $\R$ to zero yields the familiar laws: diffusion $J=-M\nabla\mu$ down gradients of the chemical potential $\mu=\delta\F/\delta c$; reactions $r=-K\mathcal A$ driven by the affinity $\mathcal A=S^{\top}\mu$; phase relaxation $\dot\eta=-Lg$ with driving force $g=\delta\F/\delta\eta$; and dislocation motion $\dot X_l=P_lB_l^{-1}P_lf_l$. Here $f_l=-\delta\F/\delta X_l$ is the total force per unit length on line $l$, which contains the Peach--Koehler force~\cite{peach1950,hirth1982} $(\sigma b_l)\times\hat{\mathbf t}_l$ from the stress $\sigma$ together with the core and binding forces, and $P_l$ projects onto the allowed glide or climb directions.

These laws differ in one important respect. A conserved quantity such as composition cannot disappear locally; it can only be moved by a flux, so $c$ changes by $\dot c=\nabla\cdot(M\nabla\mu)$ and its total is fixed (the Cahn--Hilliard equation~\cite{cahn1958}, or Fick's law for dilute species). A non-conserved quantity such as the phase field or damage can change in place, so $\eta$ relaxes locally by $\dot\eta=-Lg$ (the Allen--Cahn equation~\cite{allen1979}); \cref{fig:theory}c contrasts the two. With the energy of \cref{eq:terms}, the Allen--Cahn law moves a curved interface with a normal velocity proportional to its curvature, which is the classical law of grain-boundary motion~\cite{allen1979,mullins1956}, and the chemical potential of a curved precipitate or pore contains the Gibbs--Thomson term $\gamma/R$ for radius $R$; both are recovered quantitatively in \cref{sec:verification}.

Substituting these laws into $\dot\F$ shows that, for an isothermal closed system at mechanical equilibrium, the free energy can only decrease:
\begin{equation}\label{eq:dissipation}
\begin{aligned}
\dot\F={}&-\int_\Omega\bigl(\nabla\mu\cdot M\nabla\mu+g^{\top}Lg+\mathcal A^{\top}K\mathcal A\\
&\qquad+\mathcal{D}_{\mathrm{pl}}+\mathcal{D}_q\bigr)\,\dd V\\
&-\sum_l\int f_l\cdot P_lB_l^{-1}P_lf_l\,\dd s\;\le 0 ,
\end{aligned}
\end{equation}
where $\mathcal D_{\mathrm{pl}}\ge0$ and $\mathcal D_q\ge0$ are the plastic and damage dissipation densities that follow from $\phi_{\mathrm{pl}}$ and $\phi_q$. Every term is non-negative (the first four are quadratic forms with positive operators), so each process separately dissipates energy. In the picture of the ball rolling downhill, \cref{eq:dissipation} states that every motion, whatever its kind, lowers the landscape. This is also a practical test: if a coupled calculation ever shows the free energy of a closed system rising, the implementation is wrong. Damage may only grow, so its rate is the relaxation law cut off at zero, $\dot q=L_q\max(0,-\delta\F/\delta q)$, where $L_q$ is the damage mobility; this is the minimizer of $\R$ with the quadratic potential $\phi_q=\tfrac12\dot q^2/L_q$ under the constraint $\dot q\ge0$. A finite-strain, non-isothermal version, with multiplicative plasticity and a full heat equation, follows the same pattern; its derivation is part of the archived repository (\cref{S-app:repro}) and is not needed for the results reported here. Switching off terms recovers the standard models: Allen--Cahn~\cite{allen1979} and Cahn--Hilliard~\cite{cahn1958} dynamics, multiphase fields~\cite{steinbach1996,moelans2008}, eigenstrain elasticity~\cite{mura1987,khachaturyan1983}, crystal plasticity~\cite{roters2019}, overdamped dislocation dynamics~\cite{vdg1995,arsenlis2007}, continuum dislocation dynamics~\cite{groma1997,hochrainer2014}, diffusion with reactions and regularized fracture~\cite{francfort1998,bourdin2000,miehe2010cmame}.

\subsection{Reduced representations}

A reduced or adaptive description writes the state as a function $z=\Phi(\mathbf y)$ of fewer coordinates $\mathbf y$, for example the positions of a few dislocation lines or the values on a coarse grid. This is how cost is saved: instead of following every degree of freedom, the calculation follows a smaller set that describes the state well enough, just as a dense dislocation wall can be described by a density instead of $10^5$ positions. Minimizing the same Rayleighian over the reduced rates makes every such description dissipative, and the motion it misses, $\delta v$, has the exact cost
\begin{equation}\label{eq:omitted}
\R(\mathbf T\dot{\mathbf y})-\R(v_*)=\tfrac12\,\delta v^{\top}\mathbf R\,\delta v ,
\end{equation}
where $v_*$ is the velocity of the full model, $\mathbf T=\partial\Phi/\partial\mathbf y$ and $\mathbf R$ is the dissipation metric (derivation in \cref{S-si:derivations}). In words, a reduced description can only lose accuracy through the motion it cannot represent, and the size of that loss is known exactly. Evaluating this identity requires $v_*$, that is, a full solve, which is the origin of the first limit in \cref{sec:bounds}.

\subsection{Transfers and defect reactions}

When the grid changes, each conserved quantity is transferred by exact overlap weights. If $c_i$ is the value in an old cell $\Omega_i$ and $\Omega'_j$ is a new cell of volume $V'_j$, the new value is $c'_j=\sum_i|\Omega_i\cap\Omega'_j|\,c_i/V'_j$, where $|\Omega_i\cap\Omega'_j|$ is the volume of the overlap. This conserves totals and preserves bounds: the amount of solute that was in the old cells is exactly redistributed over the new ones. The same rule, applied to dislocations, underlies the allocator: when a group of individual dislocations is replaced by a density, each dislocation is spread over the cells around it with weights that sum to one, so that no dislocation and no Burgers vector is gained or lost (\cref{fig:schematic}c). A change is accepted only if it keeps the ledgers, keeps the required resolution of each interface and core, reproduces the fields on a common audit grid and records its energy jump.

A change of representation is bookkeeping, not physics. A physical event, such as annihilation, junction formation, absorption at an interface, nucleation or the collapse of a vacancy cluster into a loop, is different: it must conserve every inventory and charge, $Q(z^+)=Q(z^-)+{}$boundary supply, where $Q$ collects the conserved quantities and $z^-$ and $z^+$ are the states before and after the event, and its energy change is computed from the same $\F$. If a material law for an event is not available, the event is switched off and reported rather than guessed.

%% file: sections/bounds.tex
\subsection{Cost of adaptivity}\label{sec:bounds}

An adaptive time step does more than update the local fields. It also solves for the long-range stress or concentration, periodically checks itself against the full model, transfers data when the grid or representation changes, and executes physical events. Its average cost is therefore
\begin{equation}\label{eq:coststep}
C_{\mathrm{step}}=C_{\mathrm{loc}}+C_{\mathrm{glob}}+\frac{C_{\mathrm{audit}}}{k_{\mathrm{a}}}+C_{\mathrm{tr}}+C_{\mathrm{ev}} ,
\end{equation}
where $C_{\mathrm{loc}}$ is the cost of the local field update, $C_{\mathrm{glob}}$ that of the long-range solve, $C_{\mathrm{audit}}$ that of a full check performed every $k_{\mathrm a}$ steps, and $C_{\mathrm{tr}}$ and $C_{\mathrm{ev}}$ the costs of transfers and events. Adaptivity reduces only $C_{\mathrm{loc}}$. In three dimensions the number of field unknowns scales as
\begin{equation}
N_{\mathrm{dof}}\sim\frac{V}{L_c^3}+\frac{A_sw}{h^3}+\frac{L_da^2}{h^3}+\frac{N_pa^3}{h^3},
\end{equation}
for a region of volume $V$ containing interfaces of total area $A_s$ and width $w$, dislocations of total length $L_d$ with core width $a$, and $N_p$ point defects, resolved with fine spacing $h$ and coarse spacing $L_c$. This number is small when defects are sparse, but the other terms of \cref{eq:coststep} do not decrease with it. A localized defect still has a long-range field, so $C_{\mathrm{glob}}$ remains, and with explicit time integration the finest cell sets the time step everywhere. Two elementary results limit the remaining terms.

\begin{proposition}[Audit cost]\label{prop:audit}
Let a full step cost $C_{\mathrm{full}}$, and let a reduced step cost $C_{\mathrm{red}}+C_{\mathrm{audit}}$, where certifying the reduced step against the full model requires at least one evaluation of the full residual, of cost $C_{\mathrm{res}}$, per step. Then the speedup satisfies
\begin{equation}
\frac{C_{\mathrm{full}}}{C_{\mathrm{red}}+C_{\mathrm{audit}}}<\frac{1}{\chi},\qquad \chi=\frac{C_{\mathrm{res}}}{C_{\mathrm{full}}}.
\end{equation}
\end{proposition}

In words, a reduced model that is checked against the full model at every step can never be faster than the check itself. For the semi-implicit spectral schemes common in PF, a full step consists of one residual evaluation and one fast Fourier transform, so that $\chi\approx1$ and per-step certification removes any gain. Reduction can pay only when a full step costs many residual evaluations ($\chi\ll1$; our implicit finite-strain solver requires 6--35 per step), or when checks are spaced out and the error growth between them is controlled.

A natural refinement policy ranks the candidate refinements $j$ by the expected error reduction per unit cost,
\begin{equation}
S_j=\frac{\rho_j^2/\omega_j}{C_j},
\end{equation}
where $\rho_j$ is the local residual of candidate $j$, $\omega_j$ its dissipative weight and $C_j$ the cost it adds. For a quadratic Rayleighian the gain $\rho_j^2/\omega_j$ of a single candidate is exact, and the gains of several candidates add only if the candidates do not interact.

\begin{proposition}[Uniform costs]\label{prop:uniform}
If all candidates in an allocation round have the same cost $C_j=C$, ranking by $S_j$ and ranking by $\rho_j$ give the same order, and greedy selection under any budget selects the same set.
\end{proposition}

Even for additive gains, greedy selection by gain per cost is not optimal (\cref{rem:greedy}). A cost-aware policy is therefore useful only when the candidates differ genuinely in cost, for example when the choice is between different physical representations, which is the situation of the allocator in \cref{sec:allocator}.

\begin{remark}\label{rem:greedy}
With a budget of 4 and candidates (cost, gain) $=(3,5),(2,3),(2,3)$, greedy selection takes the first candidate and obtains a gain of 5, whereas the other two together give a gain of 6.
\end{remark}

The interface-focused grids used here refine whole rows and columns that cross the active region: each axis keeps the fine spacing inside a band of half-width $\Delta$ around the interface and every fourth grid line outside it.

\begin{proposition}[Cell ratio]\label{prop:cells}
Let the fine band cover a fraction $\varphi=2\Delta/L_\Omega$ of each axis of a square domain of side $L_\Omega$, with the band edges on the coarse grid. Then the focused grid has
\begin{equation}
\frac{N_a}{N_u}=\Bigl(\varphi+\frac{1-\varphi}{4}\Bigr)^2
\end{equation}
times as many cells as the uniform grid, where $N_a$ and $N_u$ are the numbers of adaptive and uniform cells, and the fine region covers an area fraction $\varphi^2$.
\end{proposition}

When the cost of a step is proportional to the number of cells, the speedup should therefore follow $N_u/N_a$ and fall to unity as the fine band fills the domain ($\varphi\to1$).

%% file: sections/methods.tex
\section{Computational details}\label{sec:methods}

All benchmarks use dimensionless parameters except the copper calculations. For every test, the quantities to be compared and their acceptance criteria were fixed before the run, and a failed criterion is reported rather than redefined; analyses added afterwards are labelled post hoc. The simulation code and benchmark scripts were written with the help of an AI coding assistant (Claude, Anthropic). Its output was checked by an automated suite of more than 360 unit and regression tests (conservation, energy decrease, restart identity and analytical solutions), by the independent references listed in \cref{tab:verification}, and by repeating the allocator study on a second machine (\cref{sec:validation}). Adaptive and uniform calculations are compared at equal physical interface width, domain, coefficients, boundary conditions, time step, tolerances and hardware, with one numerical thread. The discretizations, the test protocol, the adaptive-mesh baseline, the design of the validation study and the atomistic calculations are described in \cref{S-si:methods}.

Every calculation follows the same pattern, which can be understood without prior experience of phase-field methods. The material is divided into square cells of size $h$. Each defect is stored in the form that suits it: a phase or composition as a smooth field with a value in every cell, whose interfaces are diffuse layers of fixed physical width rather than sharp surfaces; a sparse dislocation as a particle with a position and a sign; and many dislocations together as a density per cell. At each time step, the stress of all defects is computed at once from their combined eigenstrain, every field and particle then moves down the gradient of the common free energy at a rate set by its mobility, and the totals of mass and Burgers vector are checked. The step-by-step implementation, including the discrete operators, the stress kernel, the transport scheme, a worked example of a conversion, the parameters and an algorithm for a general coupled step, is given in \cref{S-si:implementation}.

\subsection{Discrete--continuum dislocation allocator}

We consider straight edge dislocations gliding along $x$ in a periodic unit square. Dislocation $i$ lies at position $\mathbf x_i=(x_i,y_i)$ and has Burgers vector $s_ib_0\hat{\mathbf e}_x$, where $b_0$ is the Burgers vector magnitude, $s_i=\pm1$ its sign and $\hat{\mathbf e}_x$ the unit vector along $x$. It moves under the applied shear stress $\tau$ and the internal shear stress $\sigma_{xy}$ of all the other dislocations with overdamped (drag-controlled) glide,
\begin{equation}\label{eq:glide}
\dot x_i=m_d\,s_i\,b_0\bigl(\sigma_{xy}(\mathbf x_i)+\tau\bigr),
\end{equation}
where $m_d$ is the dislocation mobility and $s_ib_0(\sigma_{xy}+\tau)$ is the Peach--Koehler force per unit length. This is the dissipation law of \cref{sec:contract} restricted to glide, with dissipation rate $\sum_i\dot x_i^2/m_d\ge0$. The internal stress follows from the signed dislocation density $\alpha(\mathbf x)$ (Burgers vector per unit area) through the Airy stress function $\phi$~\cite{hirth1982}, which satisfies $\nabla^4\phi=-E'\partial_y\alpha$, with $E'=2\mu_s/(1-\nu)$ the plane-strain modulus for shear modulus $\mu_s$ and Poisson's ratio $\nu$, and $\sigma_{xy}=-\partial_x\partial_y\phi$. The stress is computed with fast Fourier transforms on a $256^2$ grid of spacing $h$, and the density is smoothed by a Gaussian core of width $1.5h$. Three methods share this stress solver. In the discrete method (D), every dislocation is a particle whose Burgers vector is spread to the grid, and whose stress is read back, with cloud-in-cell weights~\cite{hockney1988}; its cost grows with the number of dislocations $N$. In the continuum method (C), the number densities $\rho_+$ and $\rho_-$ of positive and negative dislocations are transported by the conservation laws $\partial_t\rho_\pm+\partial_x(\rho_\pm v_\pm)=0$, with the glide velocities $v_\pm=\pm m_db_0(\sigma_{xy}+\tau)$ given by \cref{eq:glide}, using the second-order upwind MUSCL scheme (Monotonic Upstream-centred Scheme for Conservation Laws)~\cite{vanleer1979} with a minmod slope limiter, which keeps the densities non-negative and free of oscillations; its cost depends only on the grid. In the allocator (H), the domain is divided into $16\times16$-cell blocks. A block holding at least $n_c$ dislocations is treated as continuum and returns to discrete only when its count falls below $n_c/2$, and the assignment is refreshed every 20 steps. The threshold has a statistical meaning: a block with $n$ dislocations has a relative sampling noise of about $n^{-1/2}$, so that for $n_c=64$ a density represents each block to about 12\%.

The switch between representations is exact. When a block becomes continuum, its particles are spread into $\rho_\pm$ with the same weights used for the stress. When it becomes discrete, each cell emits $\lfloor\rho_\pm h^2\rfloor$ particles and keeps the fractional remainder as density, and a particle that glides into a continuum block is absorbed in the same way. Neither operation changes the number of dislocations or their net Burgers vector (\cref{S-app:proofs}). The test ensemble contains a dense Gaussian cloud of $5\times10^5$ dislocations (radius 0.08, 60\% positive) and 200 isolated dislocations elsewhere; all methods use the same time step. The declared acceptance criteria were: (A1) conservation of number and Burgers vector to $10^{-12}$; (A2) agreement of H with D within 5\% in plastic strain, with all 200 isolated dislocations within half a cell of their D positions; (A3) a cost of H at most one third of that of D; and (A4) an error of C above 20\% in the sparse region.

\subsection{Allocation algorithm}\label{sec:algorithm}

\Cref{alg:allocator} summarises one run of the allocator.
\begin{itemize}
\item \textbf{State.} The dislocations are held in two representations at once: a list of particles $\{(\mathbf x_i,s_i)\}$ and the grid densities $\rho_\pm$. A block $\beta$ carries a flag $c_\beta$ that is true when it is continuum.
\item \textbf{Allocation (procedure \textsc{Allocate}).} Each block is counted as its particles plus its integrated density. A discrete block turns continuum when $n_\beta\ge n_c$, and a continuum block returns to discrete only when $n_\beta<n_c/2$, so blocks near the threshold do not switch back and forth. \textsc{Absorb} converts particles into density and \textsc{Emit} converts density back into particles.
\item \textbf{Time step (lines~\ref{alg:loop}--\ref{alg:ledger2}).} Both representations contribute to one Burgers density $\alpha$, so a single FFT solve gives the internal stress felt by particles and densities alike. Particles glide by \cref{eq:glide}, densities are transported by the upwind (MUSCL--minmod) scheme, and a particle that glides into a continuum block is absorbed immediately. After every step, the number of dislocations and their net Burgers vector are checked against the ledger (line~\ref{alg:ledger2}).
\item \textbf{Reallocation.} The allocation is refreshed every $N_r=20$ steps. The criterion uses only local counts, so the allocator never runs the full discrete model to decide where it is needed; \cref{sec:bounds} shows that such a check would cancel the saving.
\end{itemize}

\begin{algorithm}[!htb]
\caption{Discrete--continuum allocation of dislocations (method H)}\label{alg:allocator}
\footnotesize
\begin{algorithmic}[1]
\Require particles $\{(\mathbf x_i,s_i)\}$; densities $\rho_\pm$ on cells of size $h$; blocks $\beta$ with flags $c_\beta$ (true: continuum)
\Require threshold $n_c$, reallocation interval $N_r$, time step $\Delta t$
\Statex
\Procedure{Allocate}{}\Comment{choose the representation of each block}
  \For{each block $\beta$}
    \State $n_\beta\gets(\text{particles in }\beta)+h^2\sum_{\text{cells}\in\beta}(\rho_++\rho_-)$\Comment{dislocation count}
    \If{$\beta$ is discrete \textbf{and} $n_\beta\ge n_c$} \State $c_\beta\gets$ true\Comment{becomes continuum}
    \ElsIf{$\beta$ is continuum \textbf{and} $n_\beta<n_c/2$} \State $c_\beta\gets$ false\Comment{becomes discrete}
    \EndIf
  \EndFor
  \State \Call{Absorb}{particles in continuum blocks}
  \State \Call{Emit}{densities in discrete blocks}
\EndProcedure
\Statex
\Procedure{Absorb}{particles}\Comment{particles $\to$ density}
  \State add each $s_i$ to $\rho_\pm$ with cloud-in-cell weights; delete the particle
\EndProcedure
\Procedure{Emit}{cells}\Comment{density $\to$ particles}
  \State create $\lfloor\rho_\pm h^2\rfloor$ particles per cell; keep the fractional remainder in $\rho_\pm$
\EndProcedure
\Statex
\State \Call{Allocate}{}
\For{$k=1,2,\dots$}\label{alg:loop}\Comment{time steps}
  \State $\alpha\gets b_0\,[\,\mathrm{deposit}(s_i)+\rho_+-\rho_-\,]$\Comment{one Burgers density}
  \State $\sigma_{xy}\gets$ FFT stress solve of $\alpha$\Comment{one stress for both}
  \State $\mathbf x_i\gets\mathbf x_i+\Delta t\,m_d s_i b_0\,(\sigma_{xy}(\mathbf x_i)+\tau)\,\hat{\mathbf e}_x$\Comment{particles glide}
  \State transport $\rho_\pm$ with velocity $\pm m_d b_0(\sigma_{xy}+\tau)$\Comment{densities move}
  \State \Call{Absorb}{particles that entered continuum blocks}
  \If{$k \bmod N_r=0$} \Call{Allocate}{}\label{alg:realloc}
  \EndIf
  \State \textbf{check} number $N$ and net Burgers vector $\sum_i s_ib_0$ unchanged\label{alg:ledger2}\Comment{ledger}
\EndFor
\end{algorithmic}
\end{algorithm}

\subsection{Dislocation wall array in copper}\label{sec:copper-methods}

To test the allocator with material parameters, we simulated a two-dimensional idealization of a dislocation cell structure in deformed copper in physical units: an array of parallel dislocation walls, referred to below as the wall array. The elastic constants are those of polycrystalline copper, with shear modulus $\mu_s=48$~GPa and Poisson's ratio $\nu=0.324$~\cite{ledbetter1974}, and the Burgers vector of the $\tfrac12\langle110\rangle$ dislocation is $b_0=0.2556$~nm. The drag coefficient $B=50~\mu$Pa\,s lies within the range 30--80~$\mu$Pa\,s used for face-centred cubic metals in dislocation dynamics~\cite{zbib2004,kubin2013}, so that the mobility in \cref{eq:glide} is $m_d=1/B$. The applied resolved shear stress is $\tau=30$~MPa. The microstructure follows the cell structures observed in copper deformed to moderate strains~\cite{mughrabi1983}: dense dislocation walls separate nearly dislocation-free cell interiors. A periodic square of side $L_\Omega=50~\mu$m contains ten walls, 0.5~$\mu$m thick and 5~$\mu$m apart, with a dislocation density $\rho_w=2\times10^{15}$~m$^{-2}$, and cell interiors with $\rho_i=10^{12}$~m$^{-2}$. This gives about $5\times10^5$ wall dislocations and $2\times10^3$ interior dislocations. Wall dislocations have 52\% positive sign (a small excess of geometrically necessary dislocations), and interior dislocations have random signs. The stress grid has $256^2$ cells ($h=195$~nm), so the smoothing width $\ell=1.5h=0.29~\mu$m is the regularization width of this mesoscale model. The allocation blocks are $4\times4$ cells (0.78~$\mu$m), smaller than the cell size, with $n_c=64$. The time step, $\Delta t=16.4$~ps, is one quarter of the time a dislocation takes to glide one cell at the largest initial velocity, and all methods run 1000 steps. For an equal-cost comparison, particle dislocation dynamics was also run with superdislocations, in which $k$ dislocations of one sign are lumped into one, and the cost of phase-field dislocation dynamics at core resolution was estimated from the measured cost of one step (\cref{S-si:pfdd}). As a further reference framework, the cost of direct-summation discrete dislocation dynamics, in which the stress on each dislocation is summed over all others, was measured for $10^3$, $2\times10^3$ and $4\times10^3$ dislocations and extrapolated as $aN^2$ to the full ensemble. The acceptance criteria, declared before the run, were: (R1) exact conservation through every conversion; (R2) agreement of H with D within 1\% in the walls, 2\% in the cell interiors and 1\% in the mean plastic strain; (R3) a cost of H at most one fifth of that of D; and (R4) a smaller cell-interior error of H than of C.

\begin{table*}[!tb]
\centering
\caption{Recovery of established models. Errors are relative unless stated otherwise. The third column lists tests that failed or limits of the evidence; the last column gives the figures (S: Supporting Information) that show the calculations.}\label{tab:verification}
\footnotesize
\renewcommand{\arraystretch}{1.1}
\begin{tabular}{@{}>{\raggedright\arraybackslash}p{0.14\linewidth}>{\raggedright\arraybackslash}p{0.45\linewidth}>{\raggedright\arraybackslash}p{0.24\linewidth}>{\raggedright\arraybackslash}p{0.1\linewidth}@{}}
\toprule
Case & Principal evidence & Failed test or limit & Figures \\
\midrule
Allen--Cahn circle & Slope of $R^2(t)$: $-1.98$ at $h=0.5$ and $-2.0007$ on the fine reference (prediction $-2L\kappa$, with mobility $L$ and gradient coefficient $\kappa$); spatial orders 2.14/2.33, temporal 1.04/1.08; velocity--curvature law with a finite-width $\kappa_c^3$ correction ($\kappa_c$: curvature) & 1\% field test fails at $n=128$; the linear velocity--curvature tests fail because of the $\kappa_c^3$ term & \ref{S-fig:pfconv}a,b; \ref{S-fig:pfkin}a,b \\
Cahn--Hilliard coarsening & Mass drift $7.6\times10^{-13}$; 16-seed SBDF2 ensemble gives a late energy exponent $0.339\pm0.008$, stable to 0.003 under time-step halving & Individual late-time trajectories do not converge; only ensemble statistics are qualified & \ref{S-fig:pfconv}a,b; \ref{S-fig:coarsening}a \\
Grain growth & Triple-junction angle error $0.937^\circ\to0.407^\circ\to0.166^\circ$; paired $96^2$/$128^2$ ensembles agree to 5.0\% (mean area) and 9.7\% (growth slope) & Individual area histories differ by 10.8--13.8\% & \ref{S-fig:pfconv}c,d; \ref{S-fig:coarsening}b; \ref{fig:usecases}f \\
Coherent precipitates & Fixed-width shape convergence; pair forces converge to 0.028\%/0.041\% ($193^2$ vs $257^2$), with orientation-dependent attraction and repulsion & Homogeneous moduli; synthetic eigenstrains & \ref{S-fig:pfkin}d--f; \ref{fig:usecases}d \\
Point defects and sinks & Planar analytical profile error $1.49\times10^{-6}$; spherical sink strength passes 39/40 tests; closed capture/release inventory to $6.4\times10^{-13}$ & Non-monotone coarse-grid capture error at the strongest reactivity & \ref{S-fig:sinks}a; \ref{S-fig:coarsening}c; \ref{S-fig:seggrowth}a \\
Dislocations & Screw far field within 0.99\%; forces equal energy derivatives; loop velocity linear in stress with slope $m_\ell b\cos(\pi/N)$ (loop mobility $m_\ell$, $N$ polygon nodes) to $2\times10^{-15}$; Foreman $\ln(L/a)$ source law & Annihilation by exact coincidence is not a calibrated capture law & \ref{S-fig:pfkin}c; \ref{S-fig:loops}; \ref{S-fig:transfer} \\
Dislocation--interface & Absorption, transmission and reflection branches with balanced residual Burgers vectors; cross-system reflection follows $\cos2\theta$ & Sustained same-system reflection fails under the declared load & \ref{S-fig:interfaces3d}a; \ref{S-fig:linefeedback}c \\
Pores & Closed vacancy/pore inventory; interface law $V=M_I(\mu-\gamma/R)$ confirmed to 0.4\% for singly degenerate mobility & No such law for squared-degenerate mobility & \ref{S-fig:sinks}c; \ref{S-fig:seggrowth}c; \ref{fig:usecases}e \\
Coupled calculation & Point population, screw lines and a diffuse interface on one adaptive grid with reciprocal forces, exact inventory and Burgers content, and decreasing energy & Two-dimensional cross-section & \ref{S-fig:linefeedback}a; \ref{S-fig:costs2}a,b \\
\bottomrule
\end{tabular}
\end{table*}

\begin{figure*}[!tb]
\centering
\includegraphics[width=\textwidth,height=0.45\textheight,keepaspectratio]{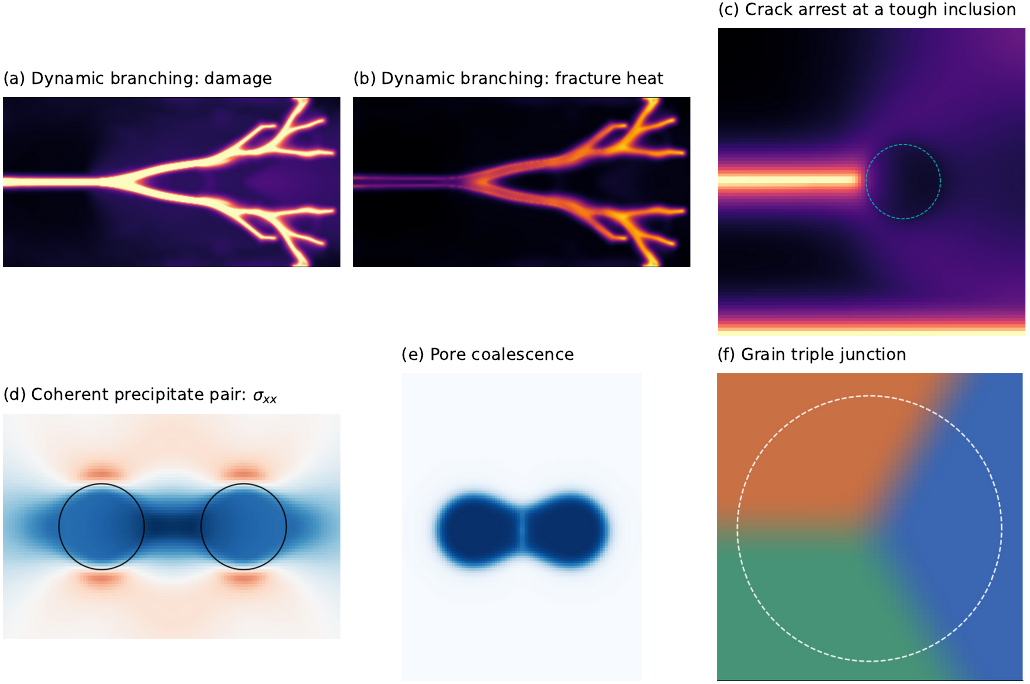}
\caption{Representative calculations within the common framework (colour scales in the source figures listed in \cref{S-app:repro}). (a, b) Dynamic crack branching in a notched $2\times1$ plate under sudden traction: damage and temperature rise at time $t=2.86$, with fracture heat and thermal softening of the toughness; total-energy drift $6.6\times10^{-5}$ of the peak. (c) Arrest of a crack at an inclusion four times tougher than the matrix. (d) Stress $\sigma_{xx}$ of two coherent precipitates with tensor eigenstrain. (e) Coalescence of two diffuse pores through their gradient flow. (f) A stationary grain triple junction at fixed physical width on a $128^2$ grid.}\label{fig:usecases}
\end{figure*}

%% file: sections/results_models.tex
\section{Results and discussion}\label{sec:results}

\subsection{Recovery of established models}\label{sec:verification}

Before any cost is compared, the framework must reproduce the established models it contains. \Cref{tab:verification} summarizes nine benchmark classes, from moving interfaces to coupled point--line--interface problems, together with the tests that failed. The semi-discrete identity \cref{S-eq:discrete} holds on every grid, and mass conservation and a non-increasing free energy were checked to round-off in every dissipative run, independently of the adaptation. \Cref{fig:usecases} gives an overview of representative calculations in which the common energy and dissipation carry several coupled mechanisms at once.

The quantitative evidence behind each entry of \cref{tab:verification} is presented in \cref{S-si:verification}: convergence and kinetic laws of the phase-field reductions (Figs.~\ref{S-fig:pfconv}--\ref{S-fig:coarsening}), vacancy sinks, segregation and pores (Figs.~\ref{S-fig:sinks} and \ref{S-fig:seggrowth}), dislocation loops, interfaces and line--field transfer (Figs.~\ref{S-fig:loops}--\ref{S-fig:linefeedback}), fracture and finite-strain mechanics (Figs.~\ref{S-fig:fracture}--\ref{S-fig:finitestrain}) and the atomistic calibration for copper (Figs.~\ref{S-fig:copper} and \ref{S-fig:clusters}). In brief, the interface motion, coarsening and grain-growth laws, the elastic interaction of precipitates, the capture and segregation of point defects, the Frank--Read and line-tension laws of dislocations, the convergence of fracture after separation and of dynamic branching, and the calibrated vacancy closure for copper all reproduce their analytical, converged or measured references within the declared tolerances; the exceptions are listed in the third column of \cref{tab:verification}.

%% file: sections/fixed_representation.tex
\subsection{Accuracy and cost of interface-focused grids}\label{sec:fixed}

We first examine the simpler form of adaptivity, in which the equations are unchanged and the grid is refined only near interfaces. \Cref{tab:fair} and \cref{S-fig:gridcost}a compare adaptive and uniform grids on the shrinking-circle Allen--Cahn benchmark and on the coupled point--line--interface problem. At $n=192$ and 256 both variants satisfy the declared accuracy tests (1\% in phase fraction, 0.05 in radius and 1\% in energy), and they differ from each other by only $4\times10^{-9}$ and $1.4\times10^{-7}$, so that adaptivity introduces no measurable error. The time-stepping speedup is 2.4--2.9, which falls to 1.2--1.5 over the whole run because start-up and output are not reduced, and the peak memory is not reduced at these sizes (\SI{84.1}{MiB} uniform against \SI{84.8}{MiB} adaptive at $n=256$). When the whole domain is active, as in spinodal decomposition, the adaptive grid coincides with the full grid and the ratio is 0.98. At $n=128$ both variants fail the field test, and that pair is not counted. For the notched fracture specimen, static grading of the mesh towards the crack path reaches the same accuracy as a uniform mesh 1.74, 4.78 and 2.56 times faster at the 10\%, 5\% and 2.5\% error levels (timed concurrently with other jobs; each ratio exceeds the 30--43\% repeat spread), with correspondingly fewer cells (\cref{S-fig:gridcost}b).

\begin{table}[!htb]
\centering
\caption{Adaptive against uniform grids at equal interface width. Ratios are uniform time over adaptive time (medians of three runs): solver (time stepping and adaptation), run (launch to exit) and DOF (uniform over adaptive number of unknowns). Error is the phase-fraction error against the finer reference.}\label{tab:fair}
\footnotesize
\setlength{\tabcolsep}{4pt}
\begin{tabular}{@{}lrrrrl@{}}
\toprule
Case & Solver & Run & DOF & Error & Tests \\
\midrule
AC, $n=128$ & 2.20 & 1.04 & 3.34 & 1.58\% & fail (both) \\
AC, $n=192$ & 2.37 & 1.18 & 3.41 & 0.68\% & pass \\
AC, $n=256$ & 2.92 & 1.52 & 3.88 & 0.29\% & pass \\
Coupled hybrid & 1.66 & 1.25 & 2.69 & -- & -- \\
Dense CH & 0.98 & 0.98 & 1.00 & identical & -- \\
\bottomrule
\end{tabular}
\end{table}

An independent quadtree adaptive-mesh solver, which matches the uniform solution to $4.3\times10^{-13}$, was 1.6--2.5 times slower than the focused grid in our Python implementations (\cref{S-si:costs}), so the comparison ranks implementations rather than adaptive mesh refinement in general.

\begin{figure}[!htb]
\centering
\includegraphics[width=\linewidth,height=0.42\textheight,keepaspectratio]{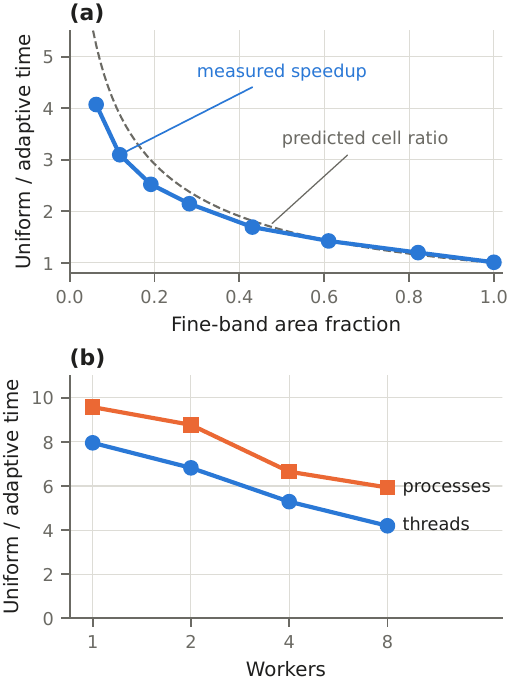}
\caption{Crossover of interface focusing. (a) Speedup over a uniform grid as the fine band grows (points), with the cell-count prediction of \cref{prop:cells} (dashed). (b) Adaptive advantage at equal numbers of threads (compiled kernel; $1024^2$ uniform against $332^2$ focused cells) and message-passing processes; all parallel results are bitwise identical to the serial ones.}\label{fig:crossover}
\end{figure}

The gain of focusing depends on how much of the domain is active (\cref{fig:crossover}a). A circle of radius $R$ between 6 and 56 on a domain of side 128 requires a fine band of half-width $R+10$. The measured cell counts agree exactly with \cref{prop:cells}, and every adaptive run matches its uniform counterpart to $3.3\times10^{-8}$. The speedup decreases steadily from 4.07 when the band covers 6\% of the domain to 2.15 at 28\% and 1.01 when it covers the whole domain. On this benchmark, interface focusing therefore yields a gain of two or more only while the fine region covers less than about 30\% of the area. The gain also decreases on parallel hardware (\cref{fig:crossover}b). With threads, the advantage over the uniform grid falls from 7.96 on one thread to 4.19 on eight, and with message-passing processes from 9.58 to 5.93. The smaller adaptive problem runs out of parallel work first (parallel efficiency 0.30 at eight threads, against 0.56 for the uniform grid), so adaptivity and parallelism partly compete for the same savings.

The limits of \cref{sec:bounds} are also visible in the measurements. Reduced models that keep a few localized coordinates for each coupled field and check themselves against the full model met their accuracy targets, yet they were 24--27 times slower than the full solver (1.26 against \SI{0.053}{s} on a $13^3$ grid and 11.69 against \SI{0.44}{s} on $29^3$), and a variant with lossless basis reset was 7--13 times slower and failed 12 of its 48 declared accuracy tests. The full step is semi-implicit and spectral, so that a check costs as much as a step ($\chi\approx1$), and \cref{prop:audit} excludes any gain from per-step certification; consistently, the slowdown did not decrease on larger grids. Likewise, a cost-weighted grid allocator produced exactly the same runs as ranking by the residual alone, as \cref{prop:uniform} requires when all candidates are equal-cost refinements.

%% file: sections/allocator.tex
\subsection{Discrete--continuum allocation of dislocations}\label{sec:allocator}

The allocator changes the representation of the dislocations themselves, which is where \cref{sec:bounds} allows a substantial gain. The discrete and continuum descriptions differ in cost roughly by the local number of dislocations, so a cost-aware choice is meaningful, and whether a region is crowded can be read from its count without a per-step check against the full model. \Cref{S-fig:allocfields} shows the plastic strain produced by the three methods for the ensemble of $5\times10^5$ densely packed and 200 isolated dislocations. The dense cloud spreads into a band of slip, while the isolated dislocations produce a faint background that the continuum method smears.

Over 83 switching events, the number of dislocations and their net Burgers vector were conserved to $1.2\times10^{-16}$ (criterion A1), and H cost \SI{2.85}{ms} per step against \SI{44.8}{ms} for D (A3). Two declared criteria failed. For A2, H matched D to 0.28\% in plastic strain, and the isolated dislocations that remained discrete matched to 0.015 cells, but nine of the 200 tracked dislocations were absorbed because the declared sparse band reached into the tail of the dense cloud, where blocks exceed $n_c$. For A4, the continuum method reached a sparse-region error of 5.5\%, much better than the expected 20\%. A longer post hoc run of 1000 steps, corresponding to about 250 cells of glide (\cref{tab:allocator}), shows that H reproduces D to 0.19\% at 17 times lower cost and, at the same cost as C, has a 2.5 times smaller error in the sparse region (0.86\% against 2.2\%). Refining C to $512^2$ increased its difference from D to 10.7\%, because D on $256^2$ depends on its own grid through the deposition; this ambiguity is examined against a finer reference in \cref{sec:validation}.

\begin{table}[!htb]
\centering
\caption{Allocator against the discrete and continuum methods (1000 steps on $256^2$, post hoc run). Time is wall-clock; errors are relative $L_2$ differences in plastic strain from D over all blocks and in the sparse region.}\label{tab:allocator}
\footnotesize
\setlength{\tabcolsep}{4pt}
\begin{tabular}{@{}lrrr@{}}
\toprule
Method & Time/step & All blocks & Sparse region \\
\midrule
D (discrete) & \SI{46.8}{ms} & reference & reference \\
C (continuum) & \SI{2.70}{ms} & 0.19\% & 2.2\% \\
H (allocator, $n_c=64$) & \SI{2.71}{ms} & 0.19\% & 0.86\% \\
\bottomrule
\end{tabular}
\end{table}

\begin{figure}[!htb]
\centering
\includegraphics[width=\linewidth]{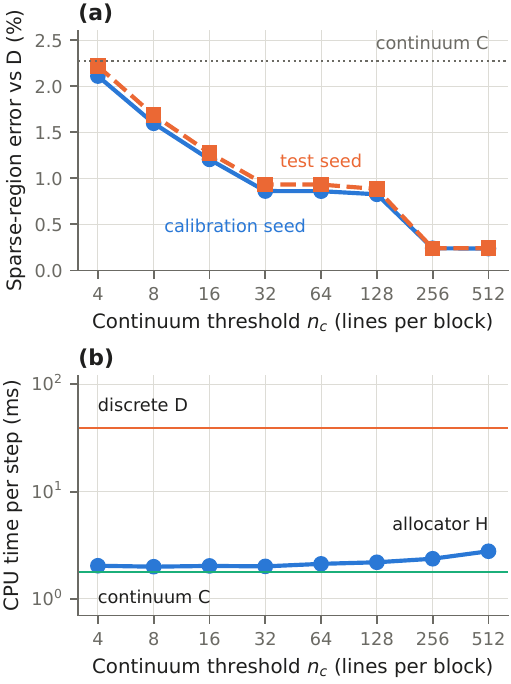}
\caption{Error control of the discrete--continuum allocator (1000 steps, $256^2$ grid, $5\times10^5$ dense and 200 isolated dislocations). (a) Sparse-region error against the discrete method D as a function of the continuum threshold $n_c$ for the calibration and test ensembles; the dotted line is the continuum method C. (b) CPU time per step of the allocator H and of the discrete and continuum methods.}\label{fig:allocator}
\end{figure}

A single threshold controls the accuracy of the allocator (\cref{fig:allocator}). We ran $n_c=4,8,\dots,512$ on two independently generated ensembles, one for calibration and one for testing. The error in the sparse region decreases steadily with $n_c$ on both ensembles, from 2.1\% when almost everything is continuum to 0.24\% when the sparse band is fully discrete (\cref{fig:allocator}a). Thresholds calibrated on the first ensemble for target errors of 2\%, 1\% and 0.5\% ($n_c=8$, 32 and 256) gave 1.69\%, 0.93\% and 0.24\% on the second, all within target, and every threshold $n_c\ge16$ was both cheaper than D and more accurate than C. The declared requirement that the cost should not decrease with $n_c$ failed because the load of the shared machine changed during the sweep; D itself ran at \SI{71}{ms} per step on one ensemble and \SI{341}{ms} on the other. Measured as CPU time, which is insensitive to other users (\cref{fig:allocator}b), D costs \SI{39.2}{ms} and C \SI{1.77}{ms} per step, and H costs between 1.99 and \SI{2.78}{ms} across the whole range of thresholds, 14--20 times cheaper than D.

%% file: sections/validation.tex
\subsection{Validation of the allocator}\label{sec:validation}

The comparison in \cref{sec:allocator} uses the discrete method on the same $256^2$ grid as its reference. Because the discrete result depends on that grid through the deposition of the Burgers vector, we repeated the study on a second computer against a reference refined in both space and time, and over a wider range of conditions (design and criteria V1--V6 in \cref{S-si:methods}). In brief, V1 tests whether the refined reference is converged, V2 whether, against it, the sparse-region error of H is within 1.25 times that of D on $256^2$ (V2a) and below that of C (V2b), V3 how the costs and errors scale with the number of dislocations, V4 how the time per step divides between the parts of the algorithm, V5 whether the errors are reproducible over five independent ensembles, and V6 whether the earlier results are reproduced on the second machine. The results are summarized in \cref{tab:validation}.

\begin{table}[!htb]
\centering
\caption{Validation of the allocator. Relative $L_2$ errors in plastic strain after 1000 steps (seed 91), over all cells and in the sparse region, against the finest discrete reference $D^*$ ($1024^2$, $\Delta t/4$; converged to 0.18\% overall but only to 6.8\% in the sparse region, so criterion V1 failed) and, for comparison, against D on $256^2$.}\label{tab:validation}
\footnotesize
\setlength{\tabcolsep}{3.5pt}
\begin{tabular}{@{}lrrrr@{}}
\toprule
 & \multicolumn{2}{c}{vs $D^*$} & \multicolumn{2}{c}{vs D ($256^2$)} \\
\cmidrule(lr){2-3}\cmidrule(l){4-5}
Method & All & Sparse & All & Sparse \\
\midrule
D ($256^2$) & 0.41\% & 15.3\% & -- & -- \\
D ($512^2$) & 0.18\% & 6.8\% & -- & -- \\
C ($256^2$) & 0.45\% & 15.5\% & 0.22\% & 2.17\% \\
C ($512^2$) & 0.23\% & 7.0\% & -- & -- \\
H ($256^2$, $n_c=64$) & 0.45\% & 15.3\% & 0.22\% & 0.86\% \\
H ($256^2$, $n_c=256$) & 0.47\% & 15.3\% & -- & -- \\
\bottomrule
\end{tabular}
\end{table}

The refined discrete calculations on $512^2$ and $1024^2$ agree to 0.18\% over all cells, so the reference is converged for the plastic strain carried by the dense population. In the sparse region, however, they still differ by 6.8\%, above the declared 2\% (criterion V1), so the reference for the isolated dislocations is converged only to that level. Against this reference, the three methods on the $256^2$ grid have nearly the same error: 15.3\% for D and H and 15.5\% for C in the sparse region, and 0.41--0.47\% overall. The allocator is therefore as accurate as the discrete method on the same grid (criterion V2a) and slightly more accurate than the continuum method (V2b), but the margin over the continuum method, which is 2.5 times in \cref{tab:allocator}, reduces to about 1\% against the refined reference. The sparse-region error of all three methods is dominated by the resolution of the shared stress grid, and it decreases to about 7\% for both D and C on $512^2$. This result clarifies what the allocator achieves. It reproduces the discrete model with which it is paired, at a fraction of its cost, but it does not remove the discretization error of the stress solver that the discrete and continuum methods share. The accuracy advantage of H over C reported in \cref{sec:allocator} is an advantage relative to the discrete model on the same grid, not relative to the grid-converged solution. \Cref{S-fig:allocfields} shows that the differences of both H and C from $D^*$ are concentrated in the dense band, where the plastic strain is largest.

The comparison with D on the same grid is robust. On five independently generated ensembles, the sparse-region error of H relative to D is 0.82--0.93\% (mean $0.86\pm0.04$\%), against 2.11--2.28\% for C (mean $2.18\pm0.06$\%), and the block-averaged error of H is 0.19\% (0.22\% over all cells) on every ensemble (criterion V5). Repeating the original measurement on the second computer reproduces the errors of 0.86\%, 2.17\% and 0.19\% reported in \cref{sec:allocator} to relative differences below $4\times10^{-14}$ (criterion V6), although the operating system, processor and Python and NumPy versions differ.

\begin{figure*}[!tb]
\centering
\includegraphics[width=0.85\textwidth]{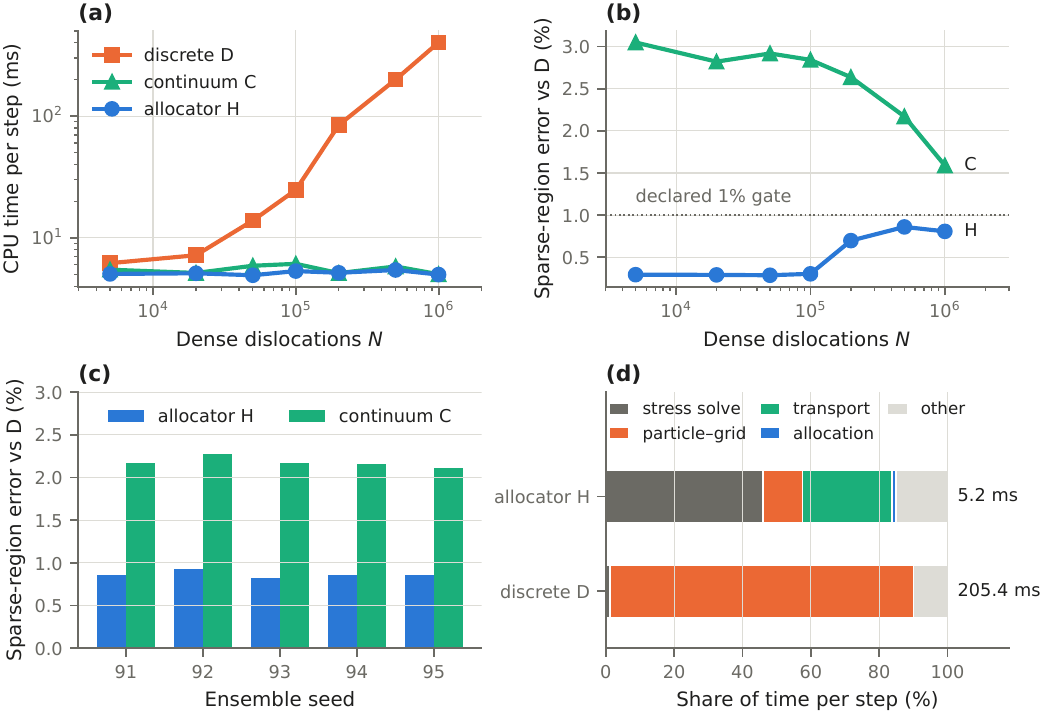}
\caption{Validation of the allocator on a second computer (1000 steps, CPU time, one thread). (a) Cost per step of the discrete (D), continuum (C) and allocator (H) methods against the number of dense dislocations $N$. (b) Sparse-region error of H and C relative to D against $N$, with the declared 1\% limit. (c) Sparse-region error of H and C relative to D for five independent ensembles ($N=5\times10^5$). (d) Share of the time per step spent in the stress solve, particle--grid transfer, density transport, allocation and other work, with the total time per step.}\label{fig:validation}
\end{figure*}

The advantage of the allocator grows with the size of the ensemble (\cref{fig:validation}a). As the number of dense dislocations $N$ increases from $5\times10^3$ to $10^6$, the cost of the discrete method rises linearly, by 406~ns per dislocation and step ($R^2=0.998$), whereas the costs of the continuum method and the allocator remain between 4.9 and 6.1~ms per step. The allocator is cheaper than the discrete method at every size, by a factor that increases from 1.2 at $N=5\times10^3$ to 37 at $N=5\times10^5$ and 81 at $N=10^6$ (criterion V3). Its error relative to the discrete method stays below 0.9\% in the sparse region at all sizes, whereas the continuum method has 1.6--3.1\% (\cref{fig:validation}b). The error of the allocator is also stable across the five ensembles (\cref{fig:validation}c).

The breakdown of the time per step explains the gain and indicates its limits (\cref{fig:validation}d). In the discrete method, 89\% of the time is spent on transferring particles to and from the grid, which scales with $N$, whereas the stress solve takes only 1.2\%. In the allocator, the stress solve (46\%), the density transport (26\%) and the particle transfers of the few remaining discrete dislocations (12\%) share a cost that no longer depends on $N$. The particle transfers here are written in NumPy; a compiled kernel would make them cheaper. If the particle work were 10 or 100 times faster while everything else stayed the same, the speedup at $N=5\times10^5$ would fall from 40, measured in the instrumented run (37 in the scaling sweep), to about 9 or 5. The allocator therefore remains advantageous with compiled particle kernels, but by a smaller factor than in our implementation, and the large factors at high $N$ are partly a consequence of the cost of particle operations in an interpreted language.

\begin{figure*}[!tb]
\centering
\includegraphics[width=0.92\textwidth,height=0.6\textheight,keepaspectratio]{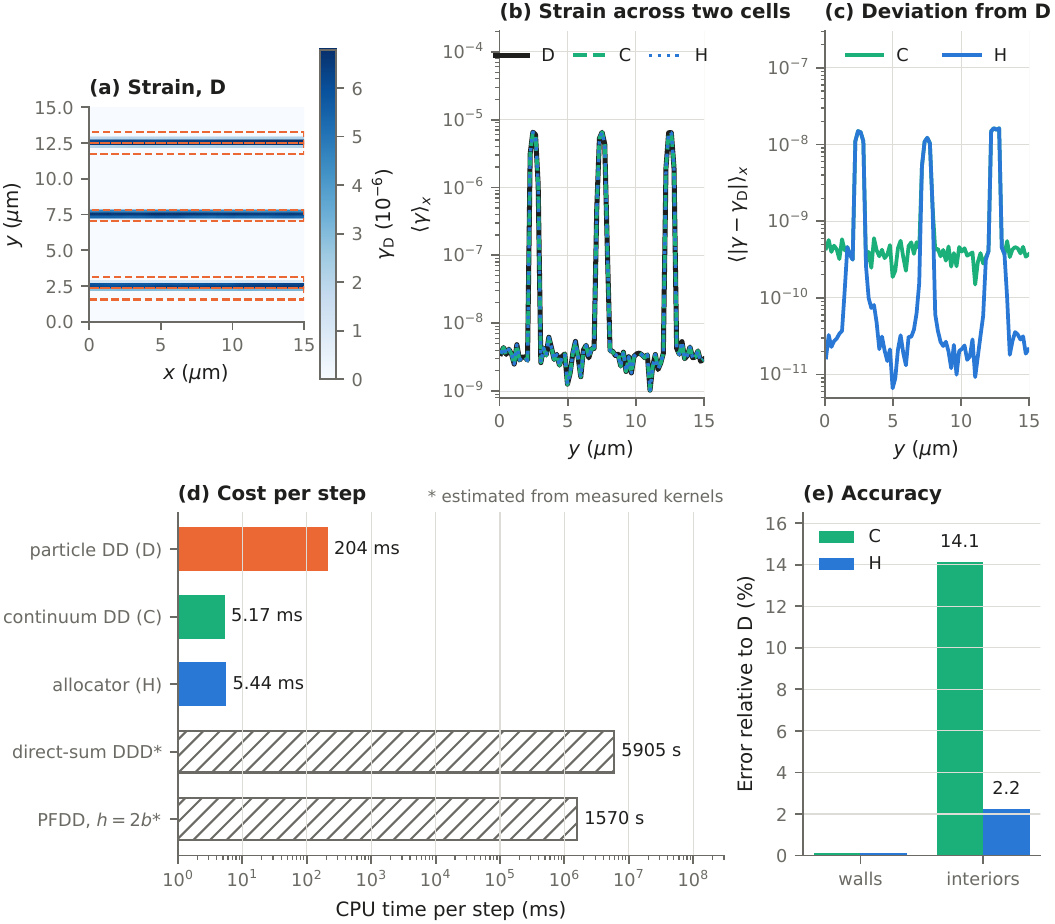}
\caption{Idealized dislocation wall array in copper ($\mu_s=48$~GPa, $b_0=0.2556$~nm, $B=50~\mu$Pa\,s, $\tau=30$~MPa; about $5\times10^5$ dislocations in a $50\times50~\mu$m$^2$ cell). (a) Plastic shear strain $\gamma$ after 1000 steps (16~ns) from particle dislocation dynamics. (b) Blocks treated as continuum by the allocator: the cell walls. (c) Deviation of the plastic strain of C and H from D, averaged along $x$, across two cells and the wall between them. (d) CPU time per step of the five frameworks; direct-summation DDD and PFDD at core resolution ($h=2b$) are estimated from measured kernels. (e) Error of C and H relative to D in the walls and in the cell interiors.}\label{fig:realistic}
\end{figure*}

%% file: sections/copper.tex
\subsection{Application to a dislocation wall array in copper}\label{sec:copper}

To demonstrate the allocator with material parameters, it was applied to an idealized wall array in copper with measured material parameters (\cref{sec:copper-methods}): about $5\times10^5$ dislocations concentrated in ten parallel walls, and about $2\times10^3$ dislocations in the interiors between them. It was compared with our implementations of particle dislocation dynamics (D) and continuum dislocation dynamics (C), and with cost estimates for direct-summation discrete dislocation dynamics and phase-field dislocation dynamics (PFDD). \Cref{tab:copper} and \cref{fig:realistic} summarize the results. After 1000 steps, the plastic strain is concentrated in the walls (\cref{fig:realistic}a). The allocator has placed the walls in continuum blocks and kept the cell interiors discrete (\cref{fig:realistic}b). At the end of the run, one quarter of the blocks are continuum, and the roughly $2\times10^3$ interior dislocations are tracked individually. The number of dislocations and their net Burgers vector are conserved exactly through every conversion (criterion R1).

\begin{table}[!htb]
\centering
\caption{Idealized dislocation wall array in copper (about $5\times10^5$ dislocations, 1000 steps, 16~ns). Cost is CPU time per step on one thread. Errors are relative $L_2$ differences in plastic strain from D. The last two rows are estimated from measured kernels.}\label{tab:copper}
\footnotesize
\setlength{\tabcolsep}{3.5pt}
\begin{tabular}{@{}lrrr@{}}
\toprule
Framework & Cost per step & Walls & Interiors \\
\midrule
Particle DD (D) & 204~ms & -- & -- \\
Continuum DD (C) & 5.2~ms & 0.13\% & 14.1\% \\
Allocator (H) & 5.4~ms & 0.13\% & 2.2\% \\
Direct-sum DDD & $\approx6\times10^{3}$~s & -- & -- \\
PFDD, $h=2b$ & $\approx1.6\times10^{3}$~s & -- & -- \\
\bottomrule
\end{tabular}
\end{table}

The allocator combines the advantages of the discrete and continuum descriptions in this microstructure. At 5.4~ms per step it is 37.5 times cheaper than particle dislocation dynamics (criterion R3) and costs almost the same as continuum dislocation dynamics. In the walls, where both H and C use the continuum description, their errors are equal (0.13\%). In the sparse cell interiors, the continuum description smears individual dislocations and has an error of 14.1\%, whereas the allocator, which tracks them as particles, has 2.2\% (criterion R4). The deviation profile across a cell (\cref{fig:realistic}c) shows the same result locally: H and C deviate equally at the walls, but inside the cells the deviation of H is about ten times smaller. The interior error of H slightly exceeds the declared 2\% (criterion R2 failed), whereas the wall error and the error of the mean plastic strain (0.09\%) are well within their limits.

The two frameworks that resolve dislocation cores are far more expensive for a system of this size (\cref{fig:realistic}d). Direct-summation discrete dislocation dynamics, whose cost grows quadratically with the number of dislocations, would need about $6\times10^3$~s per step. Phase-field dislocation dynamics must resolve each core on the grid~\cite{koslowski2002,wang2001}. With a grid spacing $h=2b$ over the $50~\mu$m cell it would need about $10^{10}$ grid points, $4\times10^{11}$~bytes of memory for five fields, and about $1.6\times10^3$~s per step, estimated from the measured cost of one PFDD step on grids of up to $4096^2$ points (\cref{S-si:pfdd}). These frameworks provide core-level detail that the present mesoscale model does not. For the collective plastic strain of a wall array, however, they are roughly $3\times10^5$ times more expensive per step than the allocator and exceed the memory of a single computer. Production DDD codes reduce the pairwise cost with fast multipole or spectral methods~\cite{arsenlis2007,kubin2013}, as the particle method D does here.

\begin{figure}[!htb]
\centering
\includegraphics[width=\linewidth]{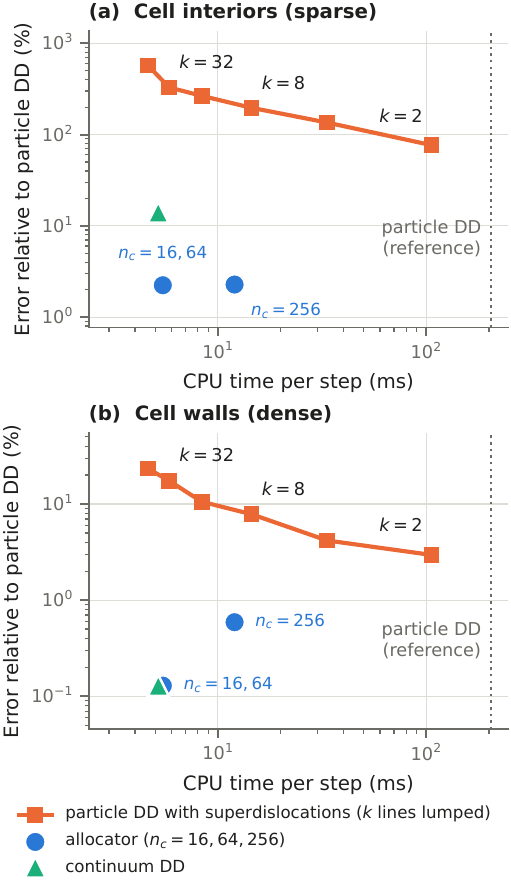}
\caption{Accuracy against cost for the copper wall array. (a) Cell interiors and (b) cell walls: error relative to particle dislocation dynamics against CPU time per step, for particle dislocation dynamics with superdislocations ($k$ dislocations lumped into one), continuum dislocation dynamics, and the allocator at three thresholds $n_c$. The dotted line marks the cost of the unlumped particle calculation.}\label{fig:pareto}
\end{figure}

A fairer test of any framework is the accuracy it reaches for a given computing budget. Particle dislocation dynamics usually handles high densities by lumping $k$ dislocations of the same sign into one superdislocation. \Cref{fig:pareto} shows the resulting trade-off. At the allocator's cost ($k=32$, 5.8~ms per step), superdislocations have an interior error of 329\% and a wall error of 18\%, against 2.2\% and 0.13\% for the allocator. Even the smallest lumping, $k=2$, at 19 times the allocator's cost, has an interior error of 77\%, because lumping destroys the sparse cell interiors. The only particle calculation at least as accurate as the allocator is the unlumped one, at 37.5 times its cost. All three declared criteria passed: at equal cost, the interior errors of superdislocations and of the continuum method are each at least five times the allocator's, and reaching the allocator's interior accuracy with particles costs at least ten times as much. In the accuracy--cost plane, the allocator therefore lies well below the curves of both reference methods for this microstructure.

Two limitations of this application should be noted. The model contains glide only, without short-range reactions or annihilation. The randomly signed wall dislocations therefore produce internal stresses of up to 0.55~GPa, which limit the time step to 16~ps and the simulated interval to 16~ns; in a real cell wall, such dislocations would form dipoles or annihilate. In addition, the $0.29~\mu$m smoothing width is the regularization width of this mesoscale model rather than the atomic core size, so the comparison concerns the collective plastic strain rather than individual core configurations.

%% file: sections/alloc3d.tex
\begin{figure*}[!tb]
\centering
\includegraphics[width=0.8\textwidth]{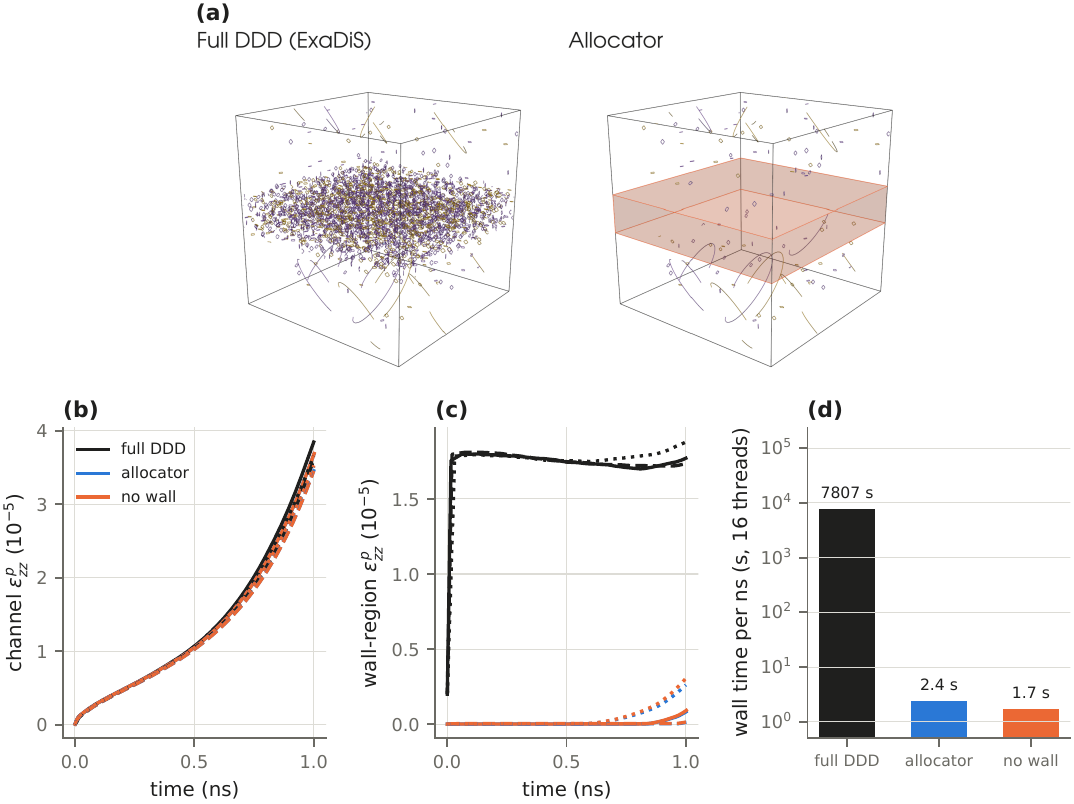}
\caption{Three-dimensional allocator test with ExaDiS in copper. (a) Networks after 1~ns (seed 1), coloured by Burgers vector: full dislocation dynamics with 3000 loops in the wall (left) and the allocator, in which the wall (shaded slab) is a continuum block acting through its friction stress (right). (b) Plastic strain swept in the channel and (c) in the wall region against time; solid, dashed and dotted lines are seeds 1, 2 and 3. (d) Wall-clock time per simulated nanosecond on 16 threads, mean over the three seeds.}\label{fig:alloc3d}
\end{figure*}

\subsection{Three-dimensional test with a production dislocation dynamics code}\label{sec:alloc3d}

The allocator was also tested in three dimensions, using ExaDiS, the high-performance engine of the OpenDiS framework~\cite{exadis,bertin2019}. ExaDiS provides both the discrete reference and the discrete part of the allocator. The test cell is a periodic copper cube $2.04~\mu$m on a side, set up as follows (details in \cref{S-si:alloc3d}):
\begin{itemize}
\item a loop-rich wall: a slab containing 3000 prismatic loops, as in the loop veins of fatigued copper~\cite{mughrabi1983};
\item a channel beside it, containing six Frank--Read sources and 100 stray loops;
\item a constant uniaxial load of 200~MPa, applied for 1~ns.
\end{itemize}
The allocator converts every block of the wall that holds at least $n_c=8$ loops into a continuum block. The removed loops leave two contributions:
\begin{itemize}
\item a stored plastic distortion, whose long-range stress on the channel (below 1\% of the applied stress) is neglected;
\item a dispersed-barrier friction stress $\tau_f=\alpha_b\mu_s b\sqrt{n_\ell d_\ell}$, with barrier strength $\alpha_b$, shear modulus $\mu_s$, loop density $n_\ell$ and loop diameter $d_\ell$, which acts on lines crossing the wall. It is entered through the spatial friction field of the ExaDiS mobility law.
\end{itemize}
The barrier strength $\alpha_b=0.1$ was chosen on seed~1 and fixed for seeds 2 and 3. As a control, a third method (no wall) deletes the wall loops and adds nothing in their place. Gates M1--M4 were declared before the runs: exact conversion (M1), channel strain within 15\% of the full calculation (M2), a no-wall error at least three times that of the allocator (M3), and a cost at most one fifth of the full calculation (M4).

\paragraph{Results} The conversion is exact. The Burgers vector removed with the loops and the error in the stored distortion are both zero to machine precision (M1). The allocator reproduces the plastic strain in the channel to within 3.9\% on the calibration seed, and to 3.3\% and 3.6\% on the two test seeds (M2, limit 15\%; \cref{fig:alloc3d}b). It is 1800--6200 times faster than the full calculation (M4), because the network shrinks from 12\,454 to about 450 segments and the time step is no longer limited by the wall loops (\cref{fig:alloc3d}d).

The control, however, is as accurate: deleting the wall gives an error of 3.4\% on both test seeds, 0.94--1.02 times those of the allocator, so M3 (no-wall error at least three times the allocator error) fails. Over 1~ns the wall hardly obstructs the sources in the channel, and the test cannot show whether the friction closure is correct. The 3.3--3.9\% difference that remains between the allocator and the full calculation comes from interactions that neither reduced method contains. The continuum wall also misses a strain component. In the full calculation the wall loops reach 90\% of their peak strain within 16--32~ps and contribute a wall-region strain of $1.8\times10^{-5}$ (\cref{fig:alloc3d}c). The frozen continuum block contains none of this strain, so the total plastic strain of the allocator is about 54\% below that of the full calculation. The test therefore confirms the exact conversion and the cost reduction in three dimensions with a production code. It does not validate the wall closure: that needs a test in which the wall controls the channel, and a continuum block that relaxes.

%% file: sections/creep.tex
\subsection{Creep cavitation and intergranular cracking in copper}\label{sec:creep}

Microstructure evolution and fracture meet in creep rupture. At elevated temperature under sustained tension, vacancies generated at grain boundaries diffuse along them to cavities, the cavities grow, and the ligaments between them finally break, producing an intergranular crack~\cite{hull1959,raj1975,kassner2003}. Classical analyses take the boundary stress as given and predict failure when the cavities coalesce~\cite{hull1959,raj1975,needleman1980}. Micromechanical studies resolve the stress redistribution but still end the process with a cohesive or coalescence criterion~\cite{onck1998,wen2018}. In the present framework, vacancy transport, cavity growth, elasticity and damage share one free energy (\cref{S-si:creep}). The load shed by the growing cavities is therefore carried by the remaining ligaments, and the ligaments fail by damage when their stress reaches the strength of the boundary.

\paragraph{Model} A grain-boundary plane in a periodic cube of copper, $10~\mu$m on a side, holds twelve cavities of initial radius 0.3--0.5~$\mu$m at random positions. The material data for copper at 773~K are taken from Frost and Ashby~\cite{frost1982}: boundary diffusivity $\delta D_b$, atomic volume $\Omega_a$ and shear modulus, together with a surface energy of 1.7~J\,m$^{-2}$~\cite{tyson1977}. Diffusion on the boundary, spherical cavities and three-dimensional elasticity with damage are solved on the GPU (\cref{S-alg:si-creep}). One closure constant, the rate at which vacancies enter a cavity, is calibrated on a single cavity against the Raj--Ashby growth law. The calibrated closure then reproduces that law to within 1\% on the calibration case and 3--8\% on four other cavity sizes and stresses, without further adjustment. Three models are compared under a constant applied stress $S$:
\begin{itemize}
\item the fixed-stress model, with the boundary stress equal to $S$ everywhere and failure at coalescence (78.5\% of the boundary cavitated), as in the classical analyses;
\item the same model with the stress redistributed around the cavities but no damage;
\item the fully coupled model, in which damage is free to separate the boundary.
\end{itemize}

\begin{figure*}[!tb]
\centering
\includegraphics[width=0.95\textwidth]{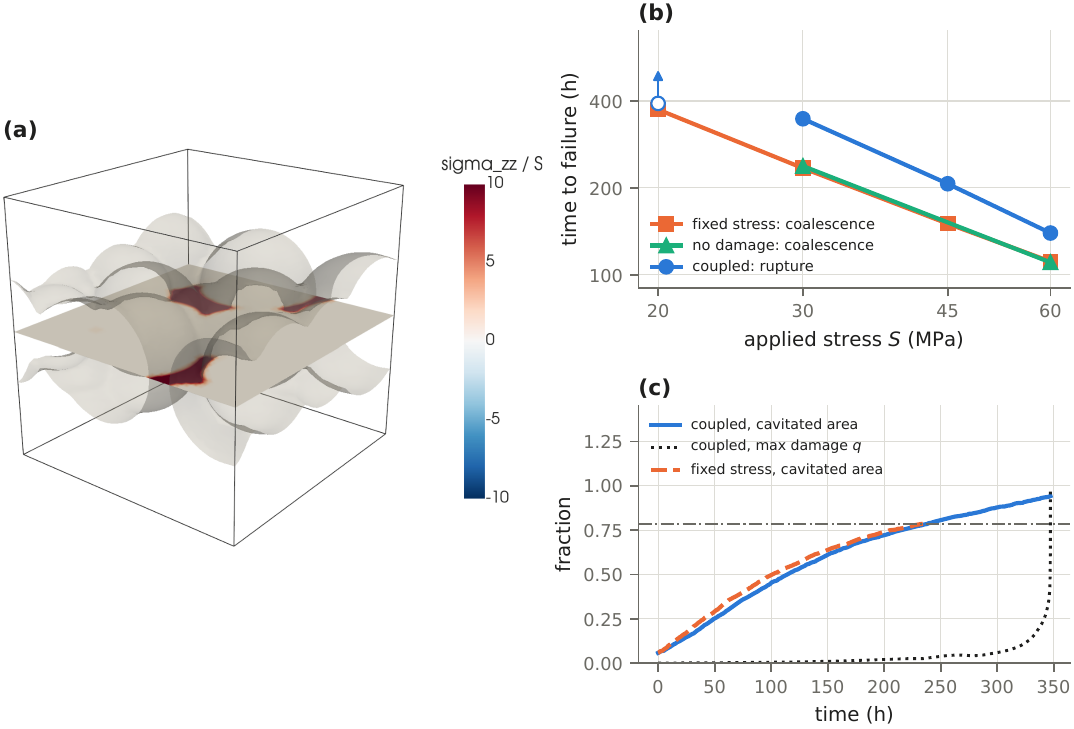}
\caption{Creep cavitation and intergranular cracking in copper at 773~K. (a) Cavities (grey surfaces) on a grain boundary in a $10~\mu$m periodic cell and the normal stress $\sigma_{zz}/S$ on the boundary at 30~MPa, just before rupture: the three remaining ligaments carry more than ten times the applied stress. (b) Time to failure against applied stress: coalescence in the fixed-stress and no-damage models, and rupture in the coupled model; the open symbol at 20~MPa is a lower bound (no rupture within 392~h). (c) Cavitated area fraction of the boundary and maximum damage against time at 30~MPa; the dash-dotted line is the coalescence criterion $\pi/4$.}\label{fig:creep3d}
\end{figure*}

\paragraph{Results} Stress redistribution alone does not change the time to coalescence: without damage the boundary reaches the coalescence criterion after 238~h at 30~MPa and 110~h at 60~MPa, the same as the fixed-stress model (234 and 111~h; \cref{fig:creep3d}b). Cavity growth is governed by boundary diffusion, which responds to the mean stress. The coupled model shows what happens after nominal coalescence. The load shed by the cavities concentrates on the remaining ligaments, which carry more than ten times the applied stress (\cref{fig:creep3d}a). These ligaments keep the boundary intact until their stress reaches the damage strength, at which point they fail within a few hours and the boundary separates (\cref{fig:creep3d}c). The resulting rupture times are 347, 207 and 140~h at 30, 45 and 60~MPa, which is 25--48\% longer than the classical coalescence estimate. At 20~MPa the boundary had not ruptured after 392~h. Without cavities, the same mechanics at 60~MPa forms no crack (maximum damage 0.003), so the rupture is produced by the interaction of cavitation and damage. Refining the grid from $48^3$ to $64^3$ changes the rupture time by 1.1\%, and changing the surface energy by $\pm20\%$ changes it by at most 3.8\%.

These results contradicted our declared expectation that stress concentration would shorten the life. Coupling instead shows that, for this cavity density and boundary strength, the classical coalescence criterion underestimates the rupture time by up to one third (rupture occurs up to 48\% later than coalescence), because the ligaments that remain at coalescence still carry load. The corresponding two-dimensional calculation, in which the boundary is a line with a lower strength, gives the opposite result: coupled failure occurs 2.0--2.3 times earlier than coalescence (\cref{S-fig:creep2d}). Which effect prevails therefore depends on the ratio of ligament strength to applied stress, which only a model coupling both processes can capture. The three-dimensional sweep was run on a $48^3$ grid after a declared $96^3$ coupled run aborted on the GPU; this change was made after the first two coupled outcomes had been seen and is reported as post hoc (\cref{S-si:creep}). The strength of the ligaments in the damage model scales as $\sqrt{EG_c/\ell}$, and only the surface energy and the grid were varied, not the damage length $\ell=0.25~\mu$m. The size of the delay therefore depends on this regularization length, and the result is a model result rather than a validated prediction of the creep life of copper.

%% file: sections/discussion.tex
\subsection{Implications for adaptive defect simulations}\label{sec:discussion}

The central result of this study is that dislocation ensembles can switch representation block by block with exact conservation and a controlled, threshold-set loss of accuracy (below 1\% against the paired discrete model). Discrete and continuum dislocation dynamics are usually alternatives chosen once for the whole simulation. In couplings such as that of Lemarchand et al.~\cite{lemarchand2001} the dislocations remain discrete and only their plastic strain is passed to a continuum, and in that of Wallin et al.~\cite{wallin2008} dislocations leave the discrete region across a fixed interface, in one direction, into conventional crystal plasticity. Here the dislocations themselves change description where the local density changes. This is possible because the conversions conserve the number and net Burgers vector of dislocations to round-off, because both descriptions feel the same stress field so that no interaction is counted twice, and because the switching decision is taken from local counts, which avoids the per-step checking cost that \cref{prop:audit} shows would cancel any gain. The allocator therefore attains the accuracy of the discrete method at close to the cost of the continuum method, with one threshold that is calibrated once and transfers between ensembles, and \cref{sec:validation} shows that its advantage grows linearly with the number of dislocations, reaching 81 times at $10^6$. Adaptive particle--continuum partitioning is established for gases, liquids and granular media~\cite{garcia1999,schwartzentruber2007,praprotnik2005,hybridgrains2018}; what this work adds is the Burgers-vector ledger and the shared energy, which make such partitioning applicable to dislocations, and the error control by a single transferable threshold, which turns the partition into a controllable approximation rather than a heuristic. The validation also delimits this contribution: against a finer discrete reference, the allocator inherits the discretization error of the stress solver it shares with the discrete method, so its contribution is to reproduce a given discrete model at a fraction of the cost, not to improve on that model's resolution.

The measurements with fixed representations indicate when grid adaptivity is worthwhile. Focusing resolution on interfaces saves approximately the fraction of cells it removes, which on our benchmark gave a gain of two or more only below about 30\% active area and no gain when the whole domain was active, as in spinodal decomposition or plasticity throughout the volume. The gain is smaller on parallel hardware, because the reduced problem saturates first. Reduced models checked at every step cannot pay when a full step is cheap, and cost-weighted refinement is equivalent to residual-based refinement when all candidates cost the same. Changing the representation, by contrast, pays where the candidate descriptions differ strongly in cost and the choice can be made in advance, as for dislocations counted block by block. Because every description in the framework shares one energy and one set of ledgers, adaptive and non-adaptive calculations solve the same model, and the measured differences in cost are not differences in physics.

The individual ingredients of the framework are established: local resolution of diffuse interfaces~\cite{provatas1999,athreya2007}, conservative transfer, regularized dislocation lines~\cite{cai2006} and variational fracture~\cite{francfort1998,bourdin2000,miehe2010cmame}. Our AMR and uniform baselines are Python implementations; compared with compiled production codes~\cite{amrex2019,prismspf2020,moose2020} the absolute rankings may change, whereas the shape of the crossover and the loss of gain on parallel hardware follow from cell counts and the work per processor and should carry over. Most calculations are two-dimensional and use dimensionless parameters, and although copper data calibrate several closures, the benchmarks are not material predictions. In the two-dimensional allocator studies the dislocations only glide, short-range reactions and annihilation are absent, and the discrete reference is a particle-in-cell model. The first three-dimensional test with the production code ExaDiS (\cref{sec:alloc3d}) confirmed exact conversion and a cost reduction of three orders of magnitude, but did not validate the continuum wall closure, because deleting the wall was equally accurate (gate M3 failed). Several declared tests failed; all are listed with their measured values and causes in \cref{S-tab:failures}, and the main ones are noted in \cref{tab:verification} and in the text. A three-dimensional test in which the wall controls the flow, with a relaxing continuum block and a comparison with production continuum dislocation dynamics~\cite{hochrainer2014,xia2015}, is the natural next step.

%% file: sections/conclusions.tex
\section{Conclusions}\label{sec:conclusions}

We have developed a variational framework in which fields, dislocation lines and defect densities share one free energy, one dissipation principle and exact ledgers of mass and Burgers vector, and used it to construct an error-controlled allocator that switches dislocations between discrete and continuum representations. The following key conclusions can be drawn from this study.

\begin{enumerate}
\item Representations can be exchanged without changing the physics: the framework reproduces the established phase-field, point-defect, dislocation and fracture models, with the exceptions listed in \cref{tab:verification}, and conserves every inventory exactly through each change of representation.
\item Switching the dislocation representation block by block, with an a priori criterion and exact Burgers-vector conservation, reproduces the discrete model at a fraction of its cost. A single threshold sets the error, and its calibration transfers between ensembles.
\item For an idealized array of dislocation walls in copper with about $5\times10^5$ dislocations, the allocator is 37.5 times cheaper than particle dislocation dynamics and six times more accurate than continuum dislocation dynamics in the cell interiors. At equal cost, particle dislocation dynamics with superdislocations is more than two orders of magnitude less accurate.
\item Because diffusion, elasticity and damage share one energy, creep cavitation and intergranular cracking can be followed to rupture. In a three-dimensional copper model at 773~K (post hoc $48^3$ sweep), the coupled model ruptures 25--48\% later than the classical coalescence estimate, because ligaments left at coalescence still carry load; in two dimensions the same coupling shortens the life. A coalescence criterion alone cannot tell which case applies.
\item Reduced models checked against the full solver at every step cannot reduce the cost, and grids refined at interfaces gain less than twofold once the active region exceeds about 30\% of the domain and nothing when it fills the domain; both limits can be predicted in advance.
\end{enumerate}

Representation switching is therefore the effective form of adaptivity when the competing descriptions differ strongly in cost and the switch can be decided locally. In three dimensions, with the production code ExaDiS, the conversion remained exact and the cost fell by three orders of magnitude, although simply deleting the wall was equally accurate, so this test establishes the exact conversion but not the value of the closure. The test wall did not constrain the channel, so the continuum closure of a three-dimensional wall is not yet validated; a test in which the wall controls the flow, with a continuum block that relaxes, is the natural next step.

%% file: sections/declarations.tex
\section*{CRediT authorship contribution statement}
\textbf{Albert Linda:} Conceptualization, Methodology, Software, Validation, Formal analysis, Investigation, Data curation, Visualization, Writing -- original draft, Writing -- review \& editing.

\section*{Declaration of competing interest}
The author declares no known competing financial interests or personal relationships that could have appeared to influence the work reported in this paper.

\section*{Funding}
This research did not receive any specific grant from funding agencies in the public, commercial, or not-for-profit sectors.

\section*{Acknowledgements}
The author thanks the developers of ExaDiS and OpenDiS (Lawrence Livermore National Laboratory) and of the NIST Interatomic Potentials Repository for making their software and potentials openly available.

\section*{Data availability}
The source code, benchmark scripts, saved numerical results, fixed random seeds, benchmark metadata and figure scripts that support the findings of this study are openly available at \url{https://github.com/albert-hzbn/discrete-continuum-dislocation-switching} (code under the MIT licence, data under CC BY 4.0), including the result directories referred to in \cref{S-app:repro}. A versioned release will be archived on Zenodo with a persistent identifier before publication.

\section*{Declaration of generative AI and AI-assisted technologies in the manuscript preparation process}
During the preparation of this work the author used an AI coding assistant (Claude, Anthropic) to help write the simulation code, run and document the benchmarks, and draft the manuscript; its use in the research itself is described in \cref{sec:methods}. After using this tool, the author reviewed and edited the content as needed and takes full responsibility for the content of the publication.